\documentclass[11pt]{article}
\usepackage{makeidx}
\usepackage{times}
\usepackage{amssymb}
\usepackage{amsmath}
\numberwithin{equation}{section}
\usepackage{setspace}
\usepackage{color}
\usepackage{graphicx}
\usepackage{rotating}
\usepackage{pdflscape}
\usepackage{epstopdf}
\usepackage[round,authoryear,comma]{natbib}
\usepackage{natbib,hyperref}
\usepackage{booktabs}
\usepackage{setspace}
\usepackage{pgfplots}
\usepackage{subcaption}
\usepackage{xcolor}
\usepackage{multirow}
\usepackage{array}

\hypersetup{
	colorlinks   = true, 
	urlcolor     = blue, 
	linkcolor    = blue, 
	citecolor   = blue 
}
\providecommand{\U}[1]{\protect\rule{.1in}{.1in}}
\newtheorem{theorem}{Theorem}[section]

\newtheorem{corollary}{Corollary}

\newtheorem{lemma}{Lemma}[section]

\parskip=\medskipamount

\newcolumntype{C}[1]{>{\centering\arraybackslash}m{#1}}

\begin{document}

\title{Pricing insurance policies with offsetting relationship \thanks{I am grateful for the comments of two anonymous reviewers, as well as the comments on earlier versions of the paper provided by Michel Denuit, Jan Dhaene, Dani\"{e}l Linders, Julien Trufin, Andr\'{e}s Villegas, participants of the \textit{KU Leuven \& Cass Business School Ph.D. Colloquium 2019} in London, participants of the \textit{International Congress in Actuarial Science and Quantitative Finance 2019} in Manizales, participants of the \textit{International Congress on Insurance: Mathematics and Economics 2019} in Munich, participants of the \textit{International Longevity Risk and Capital Markets Solutions Conference 2019} in Washington D.C., and participants of seminars at Monash University and University of New South Wales. Any remaining mistakes are mine.}}
\date{ }
	\author{Hamza Hanbali \\ Monash University, Australia \\ Email: \texttt{hamza.hanbali@monash.edu}}
\maketitle

\begin{center}
	\hrule
\end{center}
\begin{abstract}
This paper investigates the benefits of incorporating diversification effects into the pricing process of insurance policies from two different business lines. The paper shows that, for the same risk reduction, insurers pricing policies jointly can have a competitive advantage over those pricing them separately. However, the choice of competitiveness constrains the underwriting flexibility of joint pricers. The paper goes a step further by modeling explicitly the relationship between premiums and the number of customers in each line. Using the total collected premiums as a criterion to compare the competing strategies, the paper provides conditions for the optimal pricing decision based on policyholders' sensitivity to price discounts. The results are illustrated for a portfolio of annuities and assurances. Further, using non-life data from the Brazilian insurance market, an empirical exploration shows that most pairs satisfy the condition for being priced jointly, even when pairwise correlations are high.
\bigskip

\noindent
\textbf{Keywords}: Pricing decision, multiple business lines, competitiveness
\end{abstract}
\begin{center}
	\hrule
\end{center}
\newpage
\section{Introduction}
\label{Introduction}

\subsection{Motivation}
The mechanism of interest to the present paper is how diversification across different insurance business lines can be exploited to reduce prices in some them. Offsetting relationships between different product lines are often studied in the aim of reducing the risk of the combined portfolio. In the insurance industry, the combination of business lines is commonly known as natural hedging, and consists in determining optimal product mixes \citep{Kahane,CumminsNye,Grundl2006,Tsaietal2010,WangHuangYangTsai2010,WangHuangHong2013,LiHaberman2015,LucianoRegisVigna2017}. 

Natural hedging is an ex post exercises performed after the products have been launched. Exploiting the offsetting relationship between cash flows from different business lines can serve another purpose, which is price reduction. In particular, by incorporating the diversification effect ex ante in the price, joint pricers (i.e. insurers exploiting the interaction between future liabilities of the policies in the pricing) could gain a competitive advantage over stand-alone pricers (i.e. insurers relying only on individual characteristics of the policies). Specifically, offsetting effects can be anticipated by reducing the safety loadings that reflect the insurer's risk. \cite{CoxLin} find empirically that firms providing both annuities and life assurance policies tend to charge lower premiums, which supports this intuition for longevity and mortality risks. The questions that arise are (i) what are the implications of incorporating information on offsetting relationships between two policies into the pricing process? and (ii) what market conditions are most propitious for the implementation of joint pricing strategies?

\subsection{Main findings}
This paper studies joint and stand-alone pricing of insurance policies underwritten in exchange for a premium determined from an actuarial premium principle. On top of the expected present value of future payouts, a safety loading, or risk premium, is charged to reflect the insurer's risk. It is assumed that the loading is proportional to the pure premium, and that it satisfies a risk reduction constraint at portfolio level. Under this constraint, the insurer's risk reductions are equal under the two competing pricing strategies. 

As a first step, the paper analyzes the required safety loading conditionally on the proportion of policies in each business line. It is shown that there exists a \textit{competitiveness region} for joint pricers where the actual premium can be set, and that the liabilities need not be negatively correlated for this region to exist. It is also shown that there exists a \textit{critical threshold} in the proportion of the business line with the highest standard deviation per unit of expected payout (henceforth, the most risky business line), beyond which joint pricing might lead to a competitive disadvantage on the least risky business line. Additionally, it is shown that the choice of competitiveness comes with the \textit{burden of portfolio monitoring}; for a given price in the competitiveness region, a joint pricer has to maintain the proportion of policies within a specific interval in order for the actual loading to be at least equal to the required one. 

In line with earlier studies on product mix but from an ex ante perspective, the first part of the analysis draws attention to the importance of the numbers of contracts in each business line. The knowledge about these numbers is crucial in the pricing process, because they are influenced by the potential competitive advantage of the joint pricer, which is itself influenced by portfolio composition. Three elements are essential determinants of the relationship between premiums and the number of customers. The first determinant is the total market demand in each business line, which provides information on how many customers in the market are willing to buy each type of contract. The second determinant is the number of competitors in each business line, with whom the joint pricer would share the total market demand. The third determinant is the sensitivity of potential customers to the premium discount offered by the joint pricer.

Taking into account these factors, the paper goes a step further by analyzing the total collected premiums at portfolio level under each setting. It is found that the decision on the pricing strategy can be inferred from the reaction factors of policyholders, or on a related conceptual ground, from the price elasticity of demand of the joint pricer. In particular, in order for joint pricing to be rewarding in terms of total collected premiums, the riskier line has to be relatively elastic. The least risky line has to be either relatively elastic if the corresponding demand is lower than that of the riskier line, or relatively inelastic if demand is higher. The proportion of demand beyond which conditions change is related to the critical threshold found in the conditional analysis, and is also influenced by the number of competitors.

The theoretical modeling is illustrated for a portfolio of term annuities and term assurances. The results show that, as is expected, the competitiveness region exists for these two business lines, and that it is influenced by the ratio between the survival and death benefits. The results also illustrate the impact of the reaction factors on the total collected premiums.

The analysis is substantiated using loss data from the Brazilian non-life insurance market. The empirical exploration of losses from different business lines suggests that a high number of pairs satisfy the condition for being priced jointly. The importance of both individual risks and the correlation between the losses is further highlighted, thereby complementing \cite{Leland2007}'s argument on the importance of riskiness and correlation of cash flows in the context of combining financial activities. More specifically, it is found that a competitiveness region can exist even if the correlation is high. This is the case of optional auto liability and auto property damage which exhibit a correlation of 0.86, and yet, have a positive competitiveness region.

\subsection{Contribution}
The present paper adds to the literature on combining financial activities and natural hedging \citep{Grundl2006,Leland2007,Tsaietal2010,WangHuangHong2013,LiHaberman2015,LucianoRegisVigna2017,Boonen2017}. In contrast with these studies where the offsetting effect is exploited ex post, the goal here is to investigate how this effect can be incorporated ex ante in the pricing. A related work in \cite{BayraktarYoung2007} suggests that the sum of pure endowment and term assurance prices are lower when their offsetting relationship is exploited. The present paper provides further insights into the mechanisms that lead to lower premiums, and studies when such pricing strategies are favorable, as well as what challenges insurers could face if they implement them. Unlike in \cite{BayraktarYoung2007}, the safety loading dependents on portfolio composition, which provides a richer understanding of joint pricing. In particular, the results show that joint pricing leads to more favorably priced policies only under some conditions, and that factors influencing the underwriting volume must be taken into account. Further, echoing arguments on the importance of monitoring requirements and agency costs in firms' decisions to diversify \citep{Acharya2006,FigueiredoRayley}, this paper shows that the competitive advantage offered by offsetting relationships has a cost in terms of underwriting flexibility and portfolio balancing.

A related strand of the literature has focused on developing pricing models for multiple business lines under the option-pricing framework of \cite{DohertyGarven}, which incorporates the company's default risk through an insolvency put option. Contributions in this direction include the model of \cite{PhillipsCumminsAllen} that inspired a number of subsequent papers. \cite{MyersRead} study surplus allocation rules across different business lines; see \citep{DhaeneTsanakasValdezVanduffel} for a review on capital allocations. Myers and Read argued that these allocation rules can be used to determine prices, and the approach was further investigated in e.g. \cite{Zanjani2002} and \cite{SherrisvanderHoek}. \cite{GrundlSchmeiser} argue against this claim by showing that capital allocations are not required for pricing, and that they may even lead to inappropriate prices; see also \cite{Meyers} and \cite{GrundlSchmeiser2}.

\cite{Ibragimovetal2018} used the capital allocation model based on option-pricing to compare multi-line structures (or joint pricing in the present terminology) to their mono-line counter-parts (or stand-alone pricing in the present terminology). Their model suggests that multi-line risk-neutral firms would proliferate in markets with a large number of independent risks. Mono-line companies would still operate, mostly serving lines of business with asymmetric or heavy-tailed distributions, or specializing in those exhibiting high correlation with other lines. The intuition behind the results of the present paper do not conflict with those of \cite{Ibragimovetal2018}. Nevertheless, the present paper adopts a different setting, and adds to their insights at several levels. One difference with their study is that the central quantity of interest here is the price, and not the capital allocation. Besides the debate raised against pricing based on capital allocation \citep{PhillipsCumminsAllen,Meyers,GrundlSchmeiser,GrundlSchmeiser2} in which the present work does not intend to participate, focusing on price is also relevant if firms pursue growth. Recent survey studies show that the purchasing decision in some business lines, and especially in retail insurance, is influenced by prices rather than default levels \citep{Suteretal}. Another difference is that the focus here is put on the effect of contract features, portfolio composition, and market characteristics. Additionally, the analysis of the effect of demand is based on the total collected premiums, which is an important indicator of growth that shapes business decisions \citep{ZweifelEisen}. One of the additions of the present paper to their findings consists in revealing that contract features and portfolio composition are two distinct dimensions in the joint pricing decision. For instance, the diversification effect induced by negative correlation may be undermined if the overall portfolio is unbalanced with more high-risk policies. Analogously, diversification may still be beneficial even under positive correlation depending on the individual risks (as it is found for optional auto liability and auto property damage in the Brazilian market), especially if the overall portfolio is unbalanced with more low-risk policies. Another insight is that even in the absence of diversification effects, insurers may still be better off with joint pricing when their price elasticity of demand is low on the subsidizing line. 

Other studies on the impact of demand and supply on the competitive advantage of insurers can be found in \cite{Taylor}, \cite{Emms}, \cite{PantelousPassalidou} and references therein. These studies aim at deriving optimal expressions for premiums in the case of a single business line. Under comparable settings, \cite{Emms2012}, \cite{Dutang} and \cite{AsmussenChristensenThogersen} investigate properties of insurance market equilibria. The goal of the present contribution is neither to derive expressions for optimal premiums, nor to study market equilibria. Instead, the aim is to provide results which support insurance companies active on two business lines in the decision of pricing them jointly or separately.

\subsection{Structure of the paper}
The remainder of the paper is organized as follows. Section \ref{Pricing} contains some preliminary notations and the expressions of the premiums. The analysis of the required proportional loading under joint pricing is performed in Section \ref{Sec:Conditional}, conditionally on the proportion of policyholders in each business line. The section also discusses the competitiveness region, the critical threshold and the burden of portfolio monitoring. Section \ref{Sec:Unconditional} is devoted to the unconditional analysis, under a functional assumption linking policyholders' behavior and market structure to the number of policies underwritten in each business line. Section \ref{Illustration} illustrates the theoretical results for a portfolio of term annuities and term assurances. An empirical exploration of non-life business lines is performed in Section \ref{Sec:Empiricalstudy} using data from the Brazilian insurance market. This section studies the competitiveness region for the top $10$ non-life business lines in terms of gross premiums. It is found that a majority of $75.55\%$ among the total number of candidate pairs can be priced jointly. Section \ref{Conclusions} summarizes the paper and discusses possible future research. The closing section also reviews some legal limitations preventing insurers from operating in some business lines simultaneously, and discusses the particular case of bundling. All proofs are relegated to the appendix.

\section{Pricing models}
\label{Pricing}
\subsection{Preliminaries}
Consider an insurance market with two types of policies, $A$ and $B$. These policies are from two different business lines, and it is assumed that there is a single risk profile of policyholders in each business line. The per-policy present values of future payments to policyholders in each business line are denoted by $V_{A}$ and $V_{B}$, respectively. The positive and finite expected values of $V_A$ and $V_B$ are denoted by $\pi_A$ and $\pi_B$, respectively. The corresponding positive and finite standard deviations are denoted by $\sigma_A$ and $\sigma_B$. Without loss of generality, the ratio $b=\frac{\sigma_B\pi_A}{\sigma_A\pi_B}$ is assumed to satisfy $b\geq1$, i.e. $\frac{\sigma_B}{\pi_B}\geq\frac{\sigma_A}{\pi_A}$, which means that business line $B$ has a higher standard deviation per unit of average benefit. Business line $B$ is sometimes said to be riskier than business line $A$. The correlation coefficient between $V_A$ and $V_B$ is denoted by $\rho$, with $-1\leq \rho<1$; the liabilities $V_A$ and $V_B$ are not perfectly positively correlated.

The policies are underwritten in exchange for single premiums paid at policy issue. The premiums are composed of a pure premium derived from the actuarial equivalence principle, and a safety loading, or risk premium. This loading allows the insurer to compensate for the un-diversifiable part of the risk, and takes into account the fact that the actual realizations of $V_A$ and $V_B$ are likely to depart from their expected values; more on premium principles can be found in \cite{KaasGoovaertsDhaeneDenuit} and \cite{ZweifelEisen}, among others. Here, the loading is assumed to be proportional to the expected value.

Two actuarial pricing strategies are considered. The first one is when the loadings are determined separately based on the features of each contract, and this approach is referred to as \textit{stand-alone pricing}. The second one is when an insurer active on both business lines exploits their offsetting relationship by charging the same loading for both contracts, and this approach is referred to as \textit{joint pricing}. The two approaches are described in the remainder of this section.

\subsection{Stand-alone pricing}
The pure premiums correspond to the expected present values of the contracts, i.e.\ $\pi_A$ and $\pi_B$. Charging a loading allows the insurance company to reduce its exposure to a certain level. In case of stand-alone pricing, the loadings are set according to the specific risk of each contract. The loss random variables without loadings are given by $V_A-\pi_A$ and $V_B-\pi_B$. It is assumed that the insurer determines the loaded premiums such that the loss, measured by some appropriate risk measure, for each contract separately is reduced by a certain factor chosen by the insurer. This implies that the insurance company charges policyholders for (part of) the risk in the form of a safety loading.

The stand-alone loaded premiums for contracts $A$ and $B$ are denoted by $P_{A}^{\textit{sa}}$ and $P_{B}^{\textit{sa}}$, respectively, with:
$$ P_A^{\textit{sa}} = \left(1 + \psi_A\right)\pi_A, \qquad \text{ and } \qquad P_B^{\textit{sa}} = \left(1 + \psi_B\right)\pi_B,$$
where $\psi_A$ and $\psi_B$ are the loadings for each contract. Denoting by $\zeta\in(0,1)$ the risk reduction factor set by the insurer, and by $\varphi$ the risk measure, the risk reduction equations of $P_{A}^{\textit{sa}}$ and $P_{B}^{\textit{sa}}$ are written as follows:
$$ \varphi\left[V_{A} - P_{A}^{\textit{sa}}\right]=\left(1-\zeta\right)\varphi\left[V_{A} - \pi_{A}\right],$$
and
$$ \varphi\left[V_B - P_B^{\textit{sa}}\right]=\left(1-\zeta\right)\varphi\left[V_B - \pi_B\right].$$
These equations mean that the insurer's risk with the loading (i.e. $\varphi\left[V-P^{sa}\right]$) is reduced by a factor $\zeta$ compared to the case where there is no loading (i.e. $\varphi\left[V-\pi\right]$).

Different risk measures can be used in practice. For instance, the value-at-risk would allow to reduce the insurer's risk for some specific quantile level. This measure is used in practice to determine regulatory capital for financial institutions in many countries. It is however also subject to some criticisms, for instance because it is not sub-additive, i.e. the quantile of the sum is not necessarily less than the sum of the quantiles. The conditional value-at-risk, prescribed in e.g. the Swiss Solvency Test, is sub-additive. However, the conditional value-at-risk would not allow to derive explicit closed-form expressions, especially in the first part of the present analysis.

This paper uses the mean-standard deviation (MSD) risk measure, defined for some random variable $V$ as follows:
$$\varphi[V]=\pi+\gamma\sigma,\qquad \text{for }\gamma>0,$$
where $\pi$ and $\sigma$ are the expectation and standard deviation of $V$, respectively. The choice of this risk measure is justified by four arguments. First, it satisfies positive homogeneity (i.e. $\varphi[aV]=a\varphi[V]$ for $a>0$), translation invariance (i.e. $\varphi[V+a]=\varphi[V]+a$), as well as sub-additivity. Second, it is convenient to manipulate and provides deeper insights into the dynamics of the offsetting relationship between the contracts. Third, taking into account that it is sometimes used as a premium principle, the standard deviation is meaningful in the context of pricing, as in \cite{ChenChungChoiYam}; see also \cite{AsimitBoonen} for a recent applications. Fourth, as illustrated in Section \ref{Illustration} and depending on the marginal distributions of $V_A$ and $V_B$, the coefficient $\gamma$ can be tuned to match with great precision the results from other risk measures, such as the (conditional) value-at-risk. Nevertheless, this later property does not hold for all marginal distributions of $V_A$ and $V_B$.

Solving the risk reduction equations using the MSD risk measure leads to:
$$
P_{A}^{\textit{sa}} = \left(1+\zeta\gamma \frac{\sigma_{A}}{\pi_A}\right)\pi_{A}, \qquad \text{ and } \qquad P_B^{\textit{sa}} = \left(1+\zeta\gamma \frac{\sigma_{B}}{\pi_B}\right)\pi_B,
$$
and the notations $\psi_A=\zeta\gamma\frac{\sigma_A}{\pi_A}$ and $\psi_B=\zeta\gamma\frac{\sigma_B}{\pi_B}$ are introduced. Note that since $b\geq 1$, it immediately follows that $\psi_B\geq \psi_A$.

\subsection{Joint pricing}
Consider now the case where the insurer exploits the offsetting relationship between the two business lines. The notations $N_{A}$ and $N_B$, with $N_A,N_B>0$, are used for the number of underwritten policies in business lines $A$ and $B$, respectively. The proportion of policies sold in business line $B$ is denoted by $n=\frac{N_B}{N_{A}+N_B}$. Note that there are two possible assumptions regarding the set to which $N_A$ and $N_B$ belong. Namely, $N_A,N_B \in \mathbb{N}^{*}$, or $N_A,N_B \in \mathbb{R}^{*}_+$. The analysis in Section \ref{Sec:Conditional} focuses essentially on the proportion $n$, and hence, specifying the choice of the assumption is not necessary. On the other hand, in Section \ref{Sec:Unconditional}, the model used for the numbers $N_A$ and $N_B$ assumes that they belong to $\mathbb{R}^{*}_+$.

Determining the loaded premiums when the contracts are priced jointly requires some knowledge about the numbers $N_{A}$ and $N_B$. However, these quantities are unknown when the loaded premiums are set, and are likely to be impacted by the prices. Therefore, the loaded premiums in the present and following sections shall be interpreted as the conditional premiums associated to some values of $N_{A}$ and $N_B$. These conditional premiums may as well be interpreted as the required premiums associated to the value of $n$ as it unfolds. Section \ref{Sec:Unconditional} introduces further assumptions on these numbers.

Let $P_{A}^{\textit{jp}}(n)$ and $P_{B}^{\textit{jp}}(n)$ be the required loaded premiums when the insurer prices the two business lines jointly, such that:
$$
P_{A}^{\textit{jp}}(n)=\left(1+\psi(n)\right)\pi_{A}, \qquad \text{ and } \qquad P_B^{\textit{jp}}(n)=\left(1+\psi(n)\right)\pi_B,
$$
where $\psi(n)$ is the required loading proportion associated to $n$, and is the same for both contracts.

The required loaded premiums $P_{A}^{\textit{jp}}(n)$ and $P_B^{\textit{jp}}(n)$ can be determined analogously to the stand-alone case, implying that the insurer has the same risk reduction regardless of the pricing method. In particular, the loading for the combined portfolio is such that the overall loss is reduced by the same factor $\zeta$. Thus, $\psi(n)$ satisfies:
$$ \varphi\left[N_{A}\left(V_{A}-P_{A}^{\textit{jp}}(n)\right) + N_B\left(V_B-P_B^{\textit{jp}}(n)\right)\right]=(1-\zeta)\varphi\left[N_{A}\left(V_{A}-\pi_{A}\right) + N_B\left(V_B-\pi_B\right)\right].
$$
Using the MSD risk measure, the required conditional risk premium is given by:
\begin{equation}\label{Eq4-3}
	\psi(n) = \psi_A\left(\lambda_1\tilde{n}^2 - 2\lambda_2\tilde{n} + 1\right)^{\frac{1}{2}}, 
\end{equation}
where $\tilde{n} = \frac{n\pi_B}{(1-n)\pi_A+n\pi_B}$, and
$$\lambda_1 = 1 + b^2 - 2b\rho, \qquad \text{ and } \qquad \lambda_2 = 1 - b\rho.$$

The overall risk of the joint pricer is also reduced under the present setting. In particular, the difference between the overall portfolio risk when the contracts are priced separately and the overall portfolio risk when the contracts are priced jointly is:
$$(1-\zeta)\gamma\left(N_A\sigma_A+N_B\sigma_B - \sigma_{\textit{ptf}}\right),$$
where $\sigma_{\textit{ptf}}$ is the standard deviation of $N_AV_A+N_BV_B$. Since the standard deviation is sub-additive, this difference is always positive. Thus, on top of the potential competitive advantage that the insurer could achieve with the joint loading $\psi(n)$, there is also a reduction in the level of risk. In an alternative setting, it is possible to include this difference in the joint pricing, which would accentuate the potential competitive advantage of the joint pricer. Note that depending on the marginal distributions of $V_A$ and $V_B$, this observation may not hold if the MSD risk measure is replaced by the value-at-risk.

\section{Conditional analysis of joint pricing}
\label{Sec:Conditional}
This section investigates joint pricing in a conditional setting. Namely, it compares the premiums $P^{sa}$ and $P^{jp}(n)$ as a function of the proportion $n$, and hence, implicitly assumes that $n$ is given.
\subsection{Competitiveness region}
The joint pricer is said to have a competitive advantage over stand-alone pricers if the policies can be more favorably priced using the offsetting relationship between their liabilities. This would be the case if there exist values of $n\in(0,1)$ such that $\psi(n)<\psi_A$ and $\psi(n)<\psi_B$. Since $\psi(0)=\psi_A$, $\psi(1)=\psi_B$ and $b\geq1$, a \textit{competitiveness region} exists if the minimum value of $\psi(n)$ is less than $\psi_A$. 

The first question investigated in this section is whether such a competitiveness region exists. Namely, whether there exists a range of values of the proportion $n$ such that the joint loading $\psi(n)$ is smaller than the smallest stand-alone loading $\psi_A$. The following lemma provides a condition on the contracts' features such that a competitiveness region always exists, i.e.\ such that there exists $n$ satisfying $\psi(n)<\psi_A$. A proof can be found in Appendix \ref{Lemma-1-Proof}.

\begin{lemma}\label{Lemma-1}
	A competitiveness region exists for the joint pricer if and only if:
	\begin{equation}b\rho<1.\label{Cond1}\end{equation}
	Under Condition \eqref{Cond1}, the minimum loading proportion $\psi^{\min}$ is given by:
	\begin{equation}\label{Psimin}\psi^{\min} = \psi(n^{\min}) =\psi_A\sqrt{\frac{\lambda_1-\lambda_2^2}{\lambda_1}}<\psi_A,\end{equation}
	where $n^{\min}=\frac{\lambda_{2}\pi_A}{\lambda_{2}\pi_A+\left(\lambda_1-\lambda_2\right)\pi_B }\in(0,1)$, with $n^{\min}=0$ for $b\rho=1$ and $n^{\min}=1$ for $b=\rho=1$, whereas the maximum loading proportion is given by $\psi^{\max}=\psi(1)=\psi_B$.
\end{lemma}

Along the lines of previous research on optimal product mix, Lemma \ref{Lemma-1} states that there exists a unique proportion of underwritten businesses which minimizes the value of the required loaded premium. Through the lens of pricing, this result implies the existence of a competitiveness region for the joint pricer. More specifically, in order for the premium to be sufficient and for the joint pricer to gain a competitive advantage over stand-alone pricers, the actual premium for each business line should be at least equal to the minimum required loaded premium, and at most equal to the stand-alone premium. Note that for $\pi_A=\pi_B$ and $\sigma_A = \sigma_B$, it follows that $n^{\min}=\frac{1}{2}$. This means that when the contracts have identical expected values and standard deviations, the minimum required loading is reached when the portfolio is equally balanced. If in addition, the two business lines exhibit an extreme negative correlation $\rho=-1$, then it follows from the expressions of $\lambda_1$ and $\lambda_2$ that $\psi^{\min}=0$, i.e. the minimum loaded premiums correspond to the pure premiums.

A sufficient condition for the existence of the competitiveness region is a negative correlation between the liabilities of the business lines. Indeed, Condition \eqref{Cond1} is always satisfied for $\rho\leq0$. A prime example is when one business line pays a survival benefit and the other pays a death benefit. This is the typical case on which most of the literature on natural hedging has focused. Lemma \ref{Lemma-1}, however, states that the existence of the competitiveness region is not limited to contracts with negatively correlated liabilities. Joint pricing can be implemented when there is a positive correlation as well, as long as Condition \eqref{Cond1} is satisfied.

Recall that by definition, $b=\frac{\sigma_B \pi_A}{\sigma_A \pi_B}$ with $b\geq 1$, i.e. $\frac{\sigma_B}{\pi_B}\geq \frac{\sigma_A}{\pi_A}$. Condition \eqref{Cond1} gives $\frac{\sigma_B}{ \pi_B}\rho<\frac{\sigma_A}{\pi_A}$. Thus, Condition \eqref{Cond1} can be interpreted as a requirement that the diversification effect between the business lines, which is captured in the correlation $\rho$, should decrease the risk of the riskiest business line $B$ relatively to business line $A$ in order to limit the effect of subsidization across the two lines. An example where this cannot occur is when the liabilities $V_A$ and $V_B$ are highly positively correlated. In which case, the diversification effect would be too small to reduce the gap of riskiness between the two business lines. For instance, the competitiveness region does not exist for $\rho=1$.

Let $P^{\star}_A$ and $P^{\star}_B$ be the actual loaded premiums set by the joint pricer for the respective contracts, such that $P^{\star}_A=(1+\psi^{\star})\pi_A$ and $P^{\star}_B=(1+\psi^{\star})\pi_B$, with $\psi^{\star}\in\left[\psi^{\min},\psi^{\max}\right]$. In particular, $\psi^{\star}$ is a loading under joint pricing, which may be set within the competitiveness region, i.e.\ $[\psi^{min},\psi_A)$, or outside of that region, i.e.\  $[\psi_A,\psi_B]$. Suppose that the condition $b\rho<1$ is satisfied. Despite the potential competitive advantage of the joint pricer, setting a portfolio loading $\psi^{\star}$ within the competitiveness region raises two issues, to which the remainder of this section is devoted. The first one is portfolio monitoring, which originates from the convexity of the function $\psi(n)$. The second one is the existence of a critical threshold beyond which the joint pricer loses its competitive advantage on business line $A$, and this arises when $\psi_A\neq\psi_B$ as well.

\subsection{Portfolio monitoring}
One implication of the convexity of $\psi(n)$ for joint pricers is that the choice of competitiveness comes with the burden of portfolio monitoring. More specifically, in order for the required premium to be lower than or equal to the actual premium (i.e. $P^{\star}_A$ or $P^{\star}_B$), the proportion $n$ has to be maintained within the interval $[n^{\star}_{l},n^{\star}_{u}]$, where $n^{\star}_{l}$ and $n^{\star}_{u}$ are solutions of the equation:
$$ \psi\left(n\right) = \psi^{\star}.$$
Solving the second-order equation $\psi(\tilde{n})^2=\left(\psi^{\star}\right)^2$ for $\tilde{n}$, and then determining the corresponding solution for $n$, leads to:
$$
n^{\star}_l = \frac{\tilde{n}^{\star}_l\pi_A}{\tilde{n}^{\star}_l\pi_A+(1-\tilde{n}^{\star}_l)\pi_B} \qquad \text{ and } \qquad n^{\star}_u = \frac{\tilde{n}^{\star}_u\pi_A}{\tilde{n}^{\star}_u\pi_A+(1-\tilde{n}^{\star}_u)\pi_B},
$$
with
\begin{equation}\label{Eq-21}\tilde{n}^{\star}_{l} = \frac{\lambda_2}{\lambda_1} - \sqrt{ \frac{\lambda_2^2}{\lambda_1^2} - \frac{1}{\lambda_1}\left(1-\left(\frac{\psi^{\star}}{\psi_{A}}\right)^2\right)},\end{equation}
and
\begin{equation}\label{Eq-22}\tilde{n}^{\star}_{u} = \frac{\lambda_2}{\lambda_1} + \sqrt{ \frac{\lambda_2^2}{\lambda_1^2} - \frac{1}{\lambda_1}\left(1-\left(\frac{\psi^{\star}}{\psi_{A}}\right)^2\right)}.\end{equation}
In case the proportion $n$ remains within the interval $[n^{\star}_{l},n^{\star}_{u}]$, the actual loaded premiums $P^{\star}_A$ and $P^{\star}_B$ would be higher than their required counterparts $P^{\textit{jp}}_A(n)$ and $P^{\textit{jp}}_B(n)$. Otherwise, the joint pricer would be underwriting at a loss for a given risk reduction factor $\zeta$.

The flexibility of the insurer in terms of portfolio monitoring can be measured by the length of the interval $[n^{\star}_{l},n^{\star}_{u}]$, i.e. $n^{\star}_u-n^{\star}_l$. Suppose that the joint pricer sets the actual premiums equal to the lowest possible ones given by the competitiveness region, i.e. $\psi^{\star}=\psi^{\min}$. In this case, $\tilde{n}^{\star}_u=\tilde{n}^{\star}_l=\frac{\lambda_2}{\lambda_1}$, and in particular, $n^{\star}_u=n^{\star}_l=n^{\min}$. Hence, the length of this interval is $0$. This means that setting the premium equal to the lower bound of the competitiveness region, as this may arise from an analogy with previous studies on optimal product mix, is in fact the most risky choice for the insurer. More precisely, setting $\psi^{\star}=\psi^{\min}$ implies that there is only a single proportion $n$ for which the conditional loaded premium matches the actual one.

\subsection{Critical threshold}
For $\psi^{\star}\in[\psi^{\min},\psi_A)$, the joint pricer has a competitive advantage over stand-alone pricers on both business lines. For $\psi^{\star}\in(\psi_A,\psi_B)$, the joint pricer has a competitive advantage on business line $B$, and a competitive disadvantage on business line $A$. Therefore, the required loaded premium for the contract with the highest standard deviation per unit of average benefit will always be lower under joint pricing compared to its stand-alone counterpart. In contrast, for the contract with the lowest risk per unit of average benefit, there always exists a \textit{critical threshold} in the proportion $n$ beyond which the required loaded premium under joint pricing is higher than its stand-alone counterpart. In other words, beyond the critical threshold, the safer business line subsidizes the riskier one. This is formally stated in the following lemma; see Appendix \ref{Lemma-2-Proof} for a proof.

\begin{lemma}\label{Lemma-2}
	For $b> 1$, there always exists $n_{\textit{ct}}\in(0,1)$ given by:
	\begin{equation}\label{CT}n_{\textit{ct}} = \frac{2 \lambda_2\pi_A}{2 \lambda_2\pi_A + (\lambda_1-2 \lambda_2)\pi_B},\end{equation}
	such that:
	$$\left\{\begin{array}{l}
		\psi(n)< \psi_A,  \qquad \text{ for } n <n_{\textit{ct}},\\
		\psi(n)> \psi_A,  \qquad \text{ for } n >n_{\textit{ct}},
	\end{array}\right.$$
	with $\psi(n_{\textit{ct}})=\psi_A$.
\end{lemma}

The existence of the critical threshold is all the more pertinent given that the competitive advantage of the joint pricer is higher on business $B$ than on business line $A$. Indeed, denoting again by $P^{\star}_A$ and $P^{\star}_B$ the actual loaded premiums set by the joint pricer for the respective contracts, an immediate consequence of $b>1$ is that:
$$\frac{P^{\star}_A - P^{\textit{sa}}_A}{P^{\textit{sa}}_A}>\frac{P^{\star}_B - P^{\textit{sa}}_B}{P^{\textit{sa}}_B}.$$
This implies that the joint pricer could attract more policyholders on business $B$ than on business line $A$. Thus, the actual proportion $n$ may turn out to be close to $1$, and hence, beyond the critical threshold. 

A practical solution for the joint pricer to cope with the issues raised by the critical threshold consists in constraining some contract features (e.g. the benefit amounts) such that $b=1$, or equivalently, $\psi_A=\psi_B$. In this case, the joint pricer has the same competitive advantage on both business lines, leading to $n_{\textit{ct}}=1$, which means that there is no critical threshold. In practice, it may not always be possible to control all contract features. Nevertheless, insurers could still identify pairs of business lines satisfying $\psi_A\approx \psi_B$ and $b\rho<1$ without constraining contract features.

\section{Unconditional analysis of joint pricing}
\label{Sec:Unconditional}
So far, the main conclusion of the conditional analysis is that joint pricers can gain a competitive advantage over stand-alone pricers on both business lines, but this competitive advantage costs insurers their flexibility in terms of portfolio monitoring. Therefore, the number of underwritten contracts is crucial to determine the required premiums for the joint pricer.

This section seeks further insights into the pricing decision by modeling explicitly the relationship between the number of policyholders attracted in each business line and the corresponding competitive advantage. Unlike in Section \ref{Sec:Conditional}, the analysis here is not in function of the proportion $n$. In particular, the comparison between the competing strategies is performed by assuming a model for the number of policyholders. The goal is to identify conditions on market characteristics under which insurers active on both business lines can benefit from joint pricing.

\subsection{Market model}
Consider a market with $N^{T}_{A}$ and $N^{T}_{B}$ policyholders willing to buy contracts $A$ and $B$, respectively. The proportion $w^{d}=\frac{N^{T}_{B}}{N^{T}_{A}+N^{T}_{B}}$ represents the market proportion of demand for business line $B$. The market is composed of $k_{A}$ providers of contract $A$ and $k_{B}$ providers of contract $B$, with $2\leq k_{A}<N^{T}_{A}$ and $2 \leq k_{B}<N^{T}_{B}$. Among these insurers, it is assumed that there is a single insurer considering the joint pricing strategy. Thus, there are $k_{A}-1$ insurers providing contract $A$ at its stand-alone price $P^{\textit{sa}}_A$, and $k_{B}-1$ insurers providing contract $B$ at its stand-alone price $P^{\textit{sa}}_B$.

The joint pricer sets the premiums $P^{\star}_{A}$ and $P^{\star}_B$ for the respective business line, such that:
$$P^{\star}_{A}=\left(1+\psi^{\star}\right)\pi_{A}, \qquad \text{ and } \qquad P^{\star}_{B}=\left(1+\psi^{\star}\right)\pi_{B},$$
where $\psi^{\star}\in\left[\psi^{\min},\psi_B\right]$ is derived according to the risk reduction constraint introduced in Section \ref{Pricing}. The quantities $c^{\star}_{A}$ and $c^{\star}_B$ are defined as follows:
$$c^{\star}_{A}=\frac{P^{\star}_{A}}{P^{\textit{sa}}_{A}}-1, \qquad \text{ and }\qquad c^{\star}_B=\frac{P^{\star}_B}{P^{\textit{sa}}_B}-1.$$
The joint pricer is said to have a competitive advantage over stand-alone pricers on one of the business lines if the corresponding $c^{\star}$ is negative. The lower the values of $c^{\star}_{A}$ or $c^{\star}_B$, the higher the competitive advantage on the corresponding business line.

Let $N^{\star}_A$ and $N^{\star}_B$ be the numbers of policyholders attracted by the joint pricer for the loading $\psi^{\star}$. Using the MSD risk measure, and following the reasoning that leads to \eqref{Eq4-3}, the portfolio loading proportion $\psi^{\star}$ solves the equation:
\begin{equation}\label{Eq_x}\psi^{\star}=\psi_A\left(\lambda_1\tilde{n}^{\star2} - 2\lambda_2\tilde{n}^{\star} + 1\right)^{\frac{1}{2}},\end{equation}
with $\tilde{n}^{\star}=\frac{n^{\star}\pi_B}{(1-n^{\star})\pi_A+n^{\star}\pi_B}$, and $n^{\star}=\frac{N^{\star}_B}{N^{\star}_A+N^{\star}_B}$ is the proportion of policies $B$ underwritten by the insurance company active on both business lines.

The loading proportion $\psi^{\star}$ depends on the numbers of policyholders who buy each contract, which in turn depend on $\psi^{\star}$. In particular, the loading $\psi^{\star}$ appears in both sides of Equation \eqref{Eq_x}. Different functions could be applied to model this relationship. It is assumed that for $c^{\star}_{A}=c^{\star}_B=0$, the total number of policyholders is shared equally among all insurance companies. This means that in case the insurance company active on both business lines prices the contracts separately, then all companies active on $A$ underwrite $\frac{N^{T}_{A}}{k_{A}}$ contracts, and all those active on $B$ underwrite $\frac{N^{T}_{B}}{k_{B}}$ contracts.

A candidate function satisfying this property is the logistic function, such that the numbers $N^{\star}_{A}$ and $N^{\star}_{B}$ of contracts $A$ and $B$ sold by the joint pricer are given by:
$$\begin{array}{ll}N^{\star}_A &= N^T_A \left(1 - \frac{k_A-1}{k_A-1 +\exp\left(-q_Ac^{\star}_A\right)}\right),\\
	N^{\star}_B &= N^T_B \left(1 - \frac{k_B-1}{k_B-1 +\exp\left(-q_Bc^{\star}_B\right)}\right).\end{array}$$
where $q_A>0$ and $q_B>0$ are the reaction factors of policyholders in each business line. These reaction factors are related to the joint pricer's price elasticity of demand, such that a $1\%$ change in the price of a given contract is perceived by policyholders willing to buy the corresponding contract from the joint pricer as a $q\%$ change. 

The logistic function is bounded from below by $0$, and from above by the total market demand. In other words, the number of policyholders attracted by the joint pricer is at least $0$ for $\psi^{\star}\rightarrow +\infty$, and at most equal to $N^T_A$ and $N^{T}_B$ for $\psi^{\star}\rightarrow -\infty$. However, these bounds are unlikely to be be reached. Indeed, under the present setting, and provided the insurer does not underwrite at a loss, realistic values of $\psi^{\star}$ lie within the interval $\left[\psi^{\min},\psi_B\right]$, which is expected to be relatively narrow. Thus, for sufficiently small values of $c^{\star}_A$ and $c^{\star}_B$, the cumbursomness of the logistic function can be circumvented by applying a Taylor expansion, which leads to the following conveniently linear demand functions:
$$
N^{\star}_{A} = \frac{N^{T}_{A}}{k_A}\left(1 - \frac{k_{A}-1}{k_{A}}q_{A}c^{\star}_{A}\right), \qquad \text{ and } \qquad N^{\star}_{B} = \frac{N^{T}_{B}}{k_B}\left(1 - \frac{k_{B}-1}{k_{B}}q_{B}c^{\star}_{B}\right).
$$

\subsection{Pricing decision}
The decision criterion used to determine the optimal pricing strategy is the total collected premiums. In case the insurer prices the contracts jointly, the total premiums collected in each business line is $N_{A}^{\star}P^{\star}_{A}$ and $N_{B}^{\star}P^{\star}_{B}$. In case the insurer prices the contracts separately, the total market demand is shared equally among all insurers, and hence, the total premiums collected in each business line by each of the $k_A$ and $k_B$ insurers is $\frac{N^{T}_{A}}{k_A}P^{\textit{sa}}_{A}$ and $\frac{N^{T}_{B}}{k_B}P^{\textit{sa}}_B$, respectively. Therefore, the difference between the total collected premiums with and without joint pricing is given by:
$$\mathcal{D}_{\textit{ptf}} = \left(N_A^{\star}P^{\star}_A - \frac{N^{T}_A}{k_A}P^{\textit{sa}}_A\right)+\left(N_B^{\star}P^{\star}_B - \frac{N^{T}_B}{k_B}P^{\textit{sa}}_B\right).$$
Consider the proportion $\eta$ defined as follows:
\begin{equation}\label{Eta}\eta = \frac{N^T_B\frac{1}{k_B}\pi_B(\psi_B-\psi^{\min})}{N^T_A\frac{1}{k_A}\pi_A(\psi_A-\psi^{\min})+N^T_B\frac{1}{k_B}\pi_B(\psi_B-\psi^{\min})}.\end{equation}
Consider as well the critical threshold $w_{\textit{ct}}$ in the proportion of market demand for contracts $B$:
\begin{equation}\label{CriticalThreshold}		w_{\textit{ct}}=\frac{n_{\textit{ct}}}{n_{\textit{ct}} + (1-n_{\textit{ct}})\left(1+\frac{k_B-1}{k_B}\frac{\psi_B-\psi_A}{1+\psi_B}q_B\right)\frac{k_A}{k_B}},\end{equation}
where $n_{\textit{ct}}$ is the critical threshold derived from the conditional analysis in \eqref{CT}. In particular, $w_{\textit{ct}}$ corresponds to the proportion of market demand for contracts $B$ (i.e. $w^{\textit{d}}$) such that $\psi^{\star}=\psi_{A}$. It follows that $\psi^{\star}<\psi_A$ for $w^{\textit{d}}<w_{\textit{ct}}$, and $\psi^{\star}>\psi_A$ for $w^{\textit{d}}>w_{\textit{ct}}$.

The following theorem provides sufficient conditions on policyholders' reaction factors to support the decision on the pricing strategy. These conditions involve information on the market demand and supply, as well as on the contracts' features. A proof is given in Appendix \ref{Theorem-1-Proof}, which also contains the derivations for $\eta$ and $w_{\textit{ct}}$.

\begin{theorem}\label{Theorem-1}
	An insurer active on both business lines $A$ and $B$ collects more premiums at portfolio level by pricing them \emph{jointly rather than separately}, if the following conditions are satisfied:
	\begin{equation}\label{EqT1-1}q_B >\frac{k_B}{k_B-1}\frac{1+\psi_B}{1+\psi_A},\end{equation}
	and
	\begin{equation}\label{EqT1-2}\left\{\begin{array}{ll}
			(1-\eta)q_A\frac{k_A-1}{k_A}\frac{1+\psi^{\min}}{1+\psi_A}+\eta q_B\frac{k_B-1}{k_B}\frac{1+\psi^{\min}}{1+\psi_{B}}>1, & \qquad \text{for } w^d<w_{\textit{ct}},\\
			q_A  <\frac{k_A}{k_A-1}\frac{1+\psi_A}{1+\psi_B},& \qquad \text{for } w^d>w_{\textit{ct}},
		\end{array}\right.\end{equation}
	and inequality \eqref{EqT1-1} is sufficient for $w^d=w_{\textit{ct}}$.
	
	An insurer active on both business lines $A$ and $B$ collects more premiums at portfolio level by pricing them \emph{separately rather than jointly}, if the following conditions are satisfied:
	\begin{equation}\label{EqT1-3}q_B<\frac{k_B}{k_B-1},\end{equation}
	and
	\begin{equation}\label{EqT1-4}\left\{\begin{array}{ll}q_A<\frac{k_A}{k_A-1},& \qquad \text{for } w^d<w_{\textit{ct}},\\
			q_A>\frac{k_A}{k_A-1},& \qquad \text{for } w^d>w_{\textit{ct}},\end{array}\right.\end{equation}
	and inequality \eqref{EqT1-3} is sufficient for $w^d=w_{\textit{ct}}$.
\end{theorem}

Theorem \ref{Theorem-1} states that the choice of the pricing strategy can be inferred from the values of the reaction factors $q_A$ and $q_B$. In general, it it possible to determine the values of the contract features, the number of competitors $k_A$ and $k_B$, and the proportion of market demand $w^d$. Thus, the remaining task of the insurer is to estimate $q_A$ and $q_B$ based on its own experience.

These conditions imply that the insurer should reduce (resp. increase) its premiums only if the reaction of policyholders is high (resp. low) enough in order for the premium reduction (resp. increase) to result in an increase of the total premiums. In case the joint pricer does not expect to attract a sufficient number of policyholders in business line $B$, then joint pricing is not necessarily desirable. Another important implication of Theorem \ref{Theorem-1} is that even if business line $A$ subsidizes $B$ (i.e. beyond the critical threshold), joint pricing may still be rewarding for the insurer in terms of total collected premiums.

The following corollary, whose proof can be found in Appendix \ref{Corollary-proof}, provides simpler rules to support the decision on the pricing strategy.
\begin{corollary}\label{Corollary}
	An insurer active on both business lines collects more premiums at portfolio level by pricing both contracts \emph{jointly rather than separately}, if the following conditions are satisfied:
	$$q_B >2\frac{1+\psi_B}{1+\psi^{\min}},$$
	and
	$$\left\{\begin{array}{ll}
		q_A>2\frac{1+\psi_A}{1+\psi^{\min}}, & \qquad \text{for } w^d<w_{\textit{ct}},\\
		q_A<\frac{1+\psi_A}{1+\psi_B},& \qquad \text{for } w^d>w_{\textit{ct}}.
	\end{array}\right.$$
	
	An insurer active on both business lines collects more premiums at portfolio level by pricing both contracts \emph{separately rather than jointly}, if the following conditions are satisfied:
	$$q_B<1,$$
	and
	$$\left\{\begin{array}{ll}q_A<1,& \qquad \text{for } w^d<w_{\textit{ct}},\\
		q_A>2,& \qquad \text{for } w^d>w_{\textit{ct}}.\end{array}\right.$$
\end{corollary}

From Corollary \ref{Corollary}, the results of this section can be summarized as follows. Consider an insurance company active on business lines $A$ and $B$, such that $\psi_A\leq\psi_B$. Suppose that this insurance company aims at increasing its total collected premiums by pricing business lines $A$ and $B$ jointly. The company could do so if:
	\begin{itemize}
		\item business line $B$ is relatively elastic, and
		\item business line $A$ is relatively elastic if it has a lower demand than $B$, or is relatively inelastic if it has a higher demand than $B$.
	\end{itemize}

\section{Numerical illustration for annuities and life assurances}\label{Illustration}
This section illustrates the theoretical results for term annuities (business line $A$) and term assurances (business line $B$). These two business lines received substantial interest in the context of natural hedging. Early contributions include \cite{CoxLin} and \cite{Grundl2006}. A number of papers provided further insights into the natural hedging opportunities of life contingent liabilities, such as \cite{Tsaietal2010,WangHuangYangTsai2010,WangHuangHong2013,ZhuBauer} and \cite{Boonen2017}.

In what follows, the first subsection describes the random present values and the mortality model used to obtain the distribution of $V_A$ and $V_B$. The second subsection illustrates the competitiveness region under the conditional setting. The third subsection illustrates the total collected premiums and the effect of policyholders' reaction factors.

\subsection{Business lines, simulation models and data}
The random variables $V_A$ and $V_B$ are the per-policy random present values at policy issue of the annuity and assurance business lines, respectively, such that:
$$
	V_{A}=C_{A} \sum_{k=1}^{T_{A}}\text{ }_kp_x^{(A)}v(0,k),$$
and
$$
	V_{B}=C_{B} \sum_{k=1}^{T_{B}}\text{ }_{k-1}p_y^{(B)}\left(1-p_{y+k-1}^{(B)}\right)v(0,k),$$
where $T_A$ and $T_B$ are the terms of the annuity and the assurance, respectively, $C_A$ is the yearly annuity benefit payable at the end of the year, and $C_B$ is the death benefit payable at the end of the year of death. The probability $\text{ }_kp_x^{(A)}$ is the $k$-year survival probability in the term annuity business line, where all policyholders are aged $x$. The probabilities $\text{ }_{k-1}p_y^{(B)}$ and $p_{y+k-1}^{(B)}$ are the $(k-1)$-year and $1$-year survival probabilities, respectively, in the term assurance business line. In order to take into account longevity and mortality risks, all probabilities are assumed to be random variables. Market risk is not included in this analysis, and the discounting factor $v(0,k)$ is assumed to be constant, with $v(0,k)=v^k$.

The group of policyholders in the annuity business line (i.e. business line $A$) are all aged $x=60$, whereas the group of policyholders in the assurance business line (i.e. business line $B$) are all aged $y=30$. All contracts are assumed to expire after $30$ years, i.e. $T_A=T_B=30$.

The distributions of $V_A$ and $V_B$ are obtained by simulating the survival probabilities $p^{(A)}$ and $p^{(B)}$ from the Li-Lee two-population model \citep{LiLee2005}, which allows to take into account the dependence between the mortality in the two business lines. Specifically, under the Li-Lee model, the central death date $\mu^{(i)}_{x,t}$ at age $x$ and time $t$ in population $i$ is given by:
\begin{equation}
	\label{Eq3-1}
	\mu^{(i)}_{x,t} = \exp\left(\alpha_x^{(i)} + \beta_x^{(i)}\kappa_t^{(i)} + \beta_x\kappa_t\right),
\end{equation}
where $\beta_x\kappa_t$ represents the common mortality improvement for a given age $x$, and $\beta_x^{(i)}\kappa_t^{(i)}$ represents the population-specific mortality improvement for that age. The processes $\kappa_t^{(A)}$, $\kappa_t^{(B)}$ and $\kappa_t$ are simulated from correlated random walks. 

The English and Welsh population data is used for the annuity business line, and the US population data is used for the term assurance business line. Using data from two different populations allows to incorporate in the analysis the fact mortality experience in annuity and assurance businesses tend to be different. The data sets cover the period 1950-2018 and ages 30-90, and were obtained from the Human Mortality Database (\url{www.mortality.org}). The models are estimated using singular value decomposition.

\subsection{Competitiveness region}
Table \ref{TableLife} reports the values of the pure premiums per unit of benefits (i.e. $\pi_A$ and $\pi_B$ for $C_A=C_B=1$), the corresponding standard deviations, the ratio of the pure premiums and standard deviations (i.e. $\psi$ for $\zeta \gamma=1$), and the correlation coefficient $\rho$. The table shows that the business lines satisfy the condition $\psi_A<\psi_B$, i.e. $b>1$. Further, since the correlation coefficient is negative, the condition $b\rho<1$ from \eqref{Cond1} is satisfied, meaning that the competitiveness region exists.
\begin{table}[!h]\centering
	\begin{tabular}{ccc}\toprule
	& Business line $A$ & Business line $B$\\
	& Term annuities    & Term assurances  \\
	\midrule
$\pi$& $19.84$ &$0.06091786$\\
$\sigma$&$0.1821759$ &$0.004535378$\\
$\psi$ &$0.009180161$  &$0.07445072$\\
\midrule
$\rho$ & \multicolumn{2}{c}{$-0.8282$}\\
\bottomrule
\end{tabular}\caption{Values of the pure premiums per unit of benefits (i.e. $\pi_A$ and $\pi_B$ for $C_A=C_B=1$), the corresponding standard deviations, the ratio of the pure premiums and standard deviations (i.e. $\psi$ for $\zeta\gamma=1$), and the correlation coefficient $\rho$.}\label{TableLife}
\end{table}

\begin{figure}[!h]
	\centering
	\includegraphics[width=\textwidth]{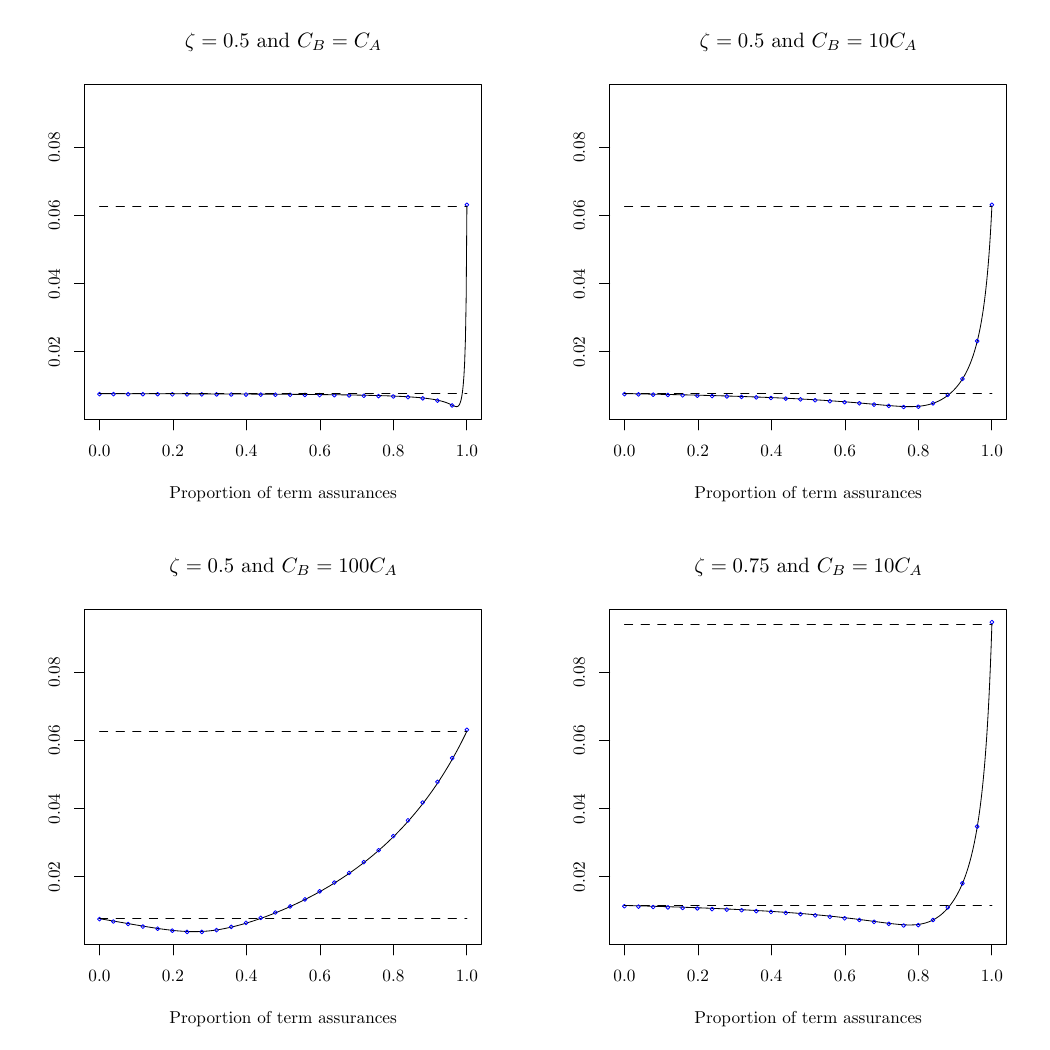}
	\caption{Premium loadings under joint pricing $\psi(n)$ in function of the proportion of term assurances $n$ for four combinations of the benefits $C_A$ and $C_B$ and the reduction factor $\zeta$. The coefficient $\gamma$ is equal to $1.686$, such that the loading $\psi(n)$ from the MSD risk measure (black curve) is approximately equal to the loading from the value-at-risk at the confidence level $0.95$ (blue circles). The horizontal dashed lines are the stand-alone loadings $\psi_A$ and $\psi_B$, with $\psi_A<\psi_B$.}
	\label{Figure1x}
\end{figure}

Figure \ref{Figure1x} displays the premium loadings under joint pricing $\psi(n)$ in function of the proportion of term assurances $n$ for four combinations of the benefits $C_A$ and $C_B$ and the reduction factor $\zeta$. The coefficient $\gamma$ is equal to $1.686$, such that the loading $\psi(n)$ from the MSD risk measure (black curve) is approximately equal to the loading obtained numerically from the value-at-risk at the confidence level $0.95$ (blue circles). Note that the black curve and the blue circles are very close, which shows the flexibility of the MSD risk measure under the present assumptions. The horizontal dashed lines are the stand-alone loadings $\psi_A$ and $\psi_B$, with $\psi_A<\psi_B$.

The results from Figure \ref{Figure1x} show how the competitiveness region is influenced by the relative levels of the benefits $C_A$ and $C_B$. In particular, the minimum loading $\psi^{\min}$ as well as the critical threshold change depending on the benefits. For instance, in case $C_B=100C_A$ (i.e. bottom-left panel), the critical threshold is lower than $0.5$, which limits the flexibility of the joint pricer. On the other hand, for $C_B=10C_A$ (i.e. top- and bottom-right panels), the critical threshold is closer to $1$, which means that setting the actual loading within the competitiveness region in this case requires less monitoring. Regarding the effect of the risk reduction factor $\zeta$, the figure shows that it is essentially a scaling factor. For instance, comparing the top- and bottom-right panels where $C_B=10C_A$ for both, the critical threshold remains unchanged. Section \ref{Sec:Empiricalstudy} contains another illustration of the competitiveness region with further discussion.

\subsection{Total collected premiums}
\begin{figure}[!h]
	\centering
 	\includegraphics[scale=0.6]{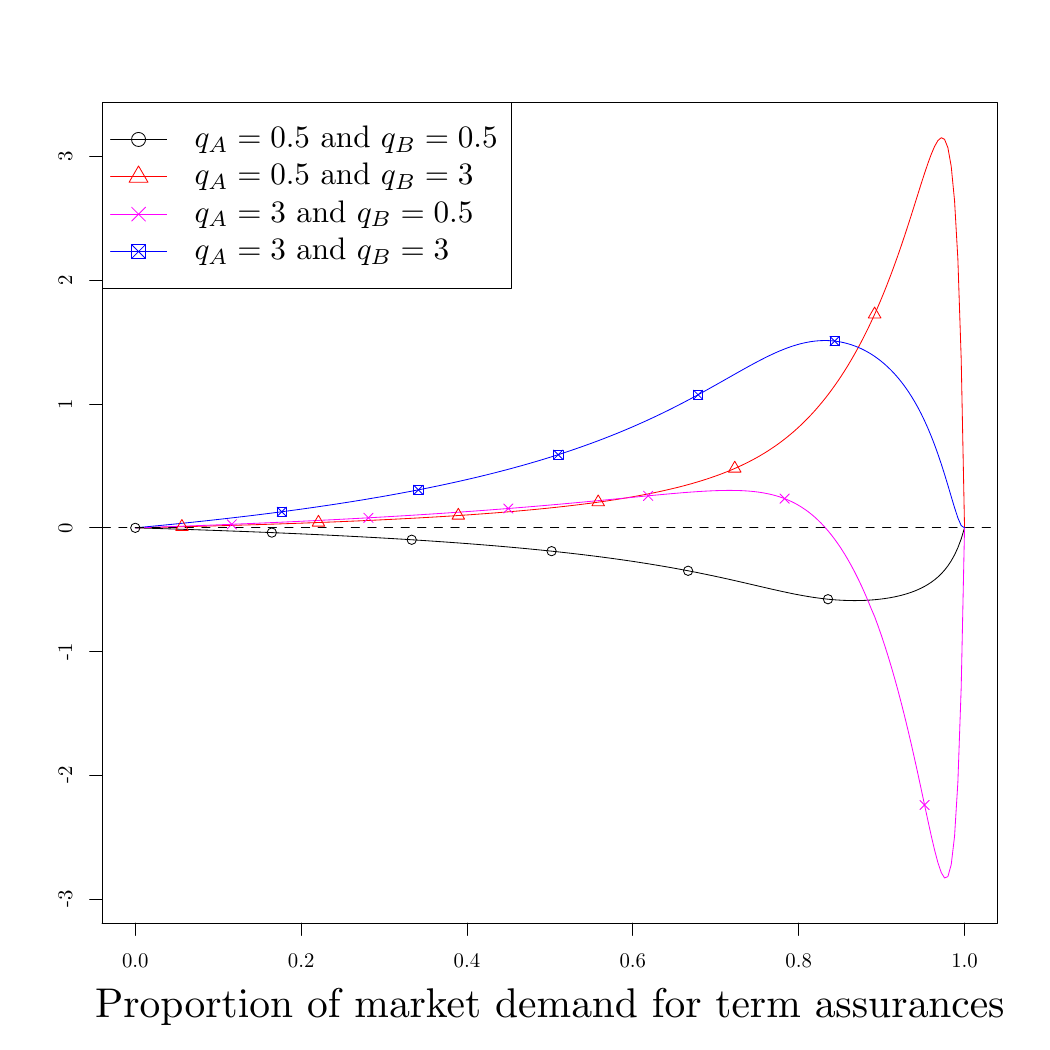}
	\caption{Relative difference of the total collected premiums with and without joint pricing in function of the proportion $w^d$ of market demand for term assurances (business line $B$) for different combinations of the reaction factors $q_A$ and $q_B$, where $C_B=10C_A$, $\zeta=0.5$, $\gamma=1.686$, and $k_A=k_B=10$. Positive (resp. negative) values indicate that joint pricing leads to higher (resp. lower) total collected premiums compared to stand-alone pricing.}
	\label{Figure2x}
\end{figure}

Figure \ref{Figure2x} displays the relative difference of the total collected premiums with and without joint pricing in function of the proportion $w^d$ of market demand for term assurances (business line $B$) for different combinations of the reaction factors $q_A$ and $q_B$, where $C_B=10C_A$, $\zeta=0.5$, $\gamma=1.686$, and $k_A=k_B=10$. Positive (resp. negative) values indicate that joint pricing leads to higher (resp. lower) total collected premiums compared to stand-alone pricing.

For $q_A=q_B=0.5$ (black curve with circles), the total collected premiums are higher under stand-alone pricing when $q_B<1$ and $q_A<1$ for $w^d<w_{ct}$, which is formally stated as a sufficient condition in Corollary \ref{Corollary}. The figure also shows that stand-alone pricing remains the preferred strategy when $q_A<1$ for $w^d\geq w_{ct}$, although it is not formally stated as a sufficient condition in the corollary. For $q_A=q_B=3$ (blue curve with squares), the opposite holds. Note that $2\frac{1+\psi_B}{1+\psi^{\min}}\approx 2.12$ and $2\frac{1+\psi_A}{1+\psi^{\min}}=2.01$, which means that the conditions of Corollary \ref{Corollary} under which joint pricing is more favorable are satisfied for $w^d<w_{ct}$. Again, even though $\frac{1+\psi_A}{1+\psi_B} \approx 0.95$, and hence $q_A>\frac{1+\psi_A}{1+\psi_B}$ for $w^d>w_{ct}$, joint pricing still leads to higher total collected premiums. A similar conclusion holds for $q_A=0.5$ and $q_B=3$ (red curve with triangles), where joint pricing leads to higher collected premiums for all $w^d$. The case where $q_A=3$ and $q_B=0.5$ (magenta curve with stars) is the only one in Figure \ref{Figure2x} where the choice of the pricing strategy depends on the critical threshold. In particular, for low proportions of market demand for term assurances, $q_A=3$ and $q_B=0.5$ leads to higher total collected premiums under joint pricing, whereas for high proportions of market demand for term assurances, stand-alone pricing leads to higher total collected premiums. Note that the latter case is stated as a sufficient condition in Corollary \ref{Corollary}.

\section{Empirical exploration of the competitiveness region for non-life business lines}\label{Sec:Empiricalstudy}
This section explores the existence of the competitiveness region for pairs of non-life business lines using a real insurance data set from the Brazilian insurance market. The first part describes the data sets. The second part explores different business lines in order to identify candidates for joint pricing. The fourth part illustrates the competitiveness region.

\subsection{Data description}

The Superintendence of Private Insurers (SUSEP), which is the executive arm of the national Brazilian insurance regulator, provides the SES data set freely accessible online (see \url{www.susep.gov.br}). It consists in aggregate losses for different business lines obtained from SES at different frequencies. The frequency used in this analysis is bi-annual. The data covers the period from July/December 2006 to July/December 2019, with 27 observations per business line. 

The raw data contains information for up to 154 business lines depending on the time period. Only non-life lines are studied. Further, business lines that were introduced or removed during the observation period were discarded. Among the remaining lines, only the top 10 business lines in terms of total collected premiums were selected. Table \ref{Table1} lists the selected lines, together with their rank in terms of total collected premiums. The advantage of working with this relatively small subset of business lines is that it allows for an easier reporting. Indeed, the total number of possible pairs of business lines is $45$.

\begin{table}[!h]
	\centering
	\begin{tabular}{p{1cm}  p{10cm} p{1cm} }
		\toprule   \centering Id. & \centering
		Business line 
		&Rank
		\\
		\midrule
		0588&Motor third party liability&4\\
		0114&Comprehensive residential& 6 \\
		0982&Private passengers auto&5 \\
		0977&Lender (except home and rural)&2\\
		0621&National transport&8\\
		1068&Mortgage (excluding Housing Finance System)&10\\
		0118& Commercial multiple peril&7\\
		
		0654&Carrier& 9 \\
		
		0553&Optional Auto Liability (in excess of mandatory cover) & 3\\	
		0531&Auto property damage& 1\\
		\bottomrule	
	\end{tabular}
	\caption{Business lines ordered based on the ratio $\psi=\sigma/\pi$, where the business line on the first row has the highest $\psi$. The first column corresponds to the identification code attributed by SUSEP. The second column contains the name of the business line translated to English from SUSEP website. The third column contains the rank of the business line in terms of total collected premiums at market level, where `1' is the highest.}	\label{Table1}
\end{table}

All time series were de-trended, after which all passed the KPSS test of stationarity \citep{KPSS}\footnote{All calculations were performed using \texttt{R} software \citep{R}, and using the packages \texttt{tseries} for times series processing and \texttt{tikzDevice} for the graphical output \citep{tseries,tikzD}}. For each pair of business lines, Condition \eqref{Cond1} is assessed to determine whether a competitiveness region exists. The random variable $V$ stands for the aggregate losses, from which the expected values (i.e. $\pi$) and the standard deviations (i.e. $\sigma$) of the individual series are calculated to obtain the ratios $\psi=\frac{\sigma}{\pi}$. The pairwise correlations $\rho$ are also calculated. For $A$ and $B$ such that $\psi_A\leq \psi_B$, if $b\rho<1$, where $b=\frac{\psi_B}{\psi_A}$, the competitiveness region of these two contracts is said to exist, indicating that insurers active on these two business lines can potentially gain a competitive advantage by pricing them jointly. 
\subsection{Analysis of business lines}
\begin{figure}[t]
	\begin{center}
		\includegraphics{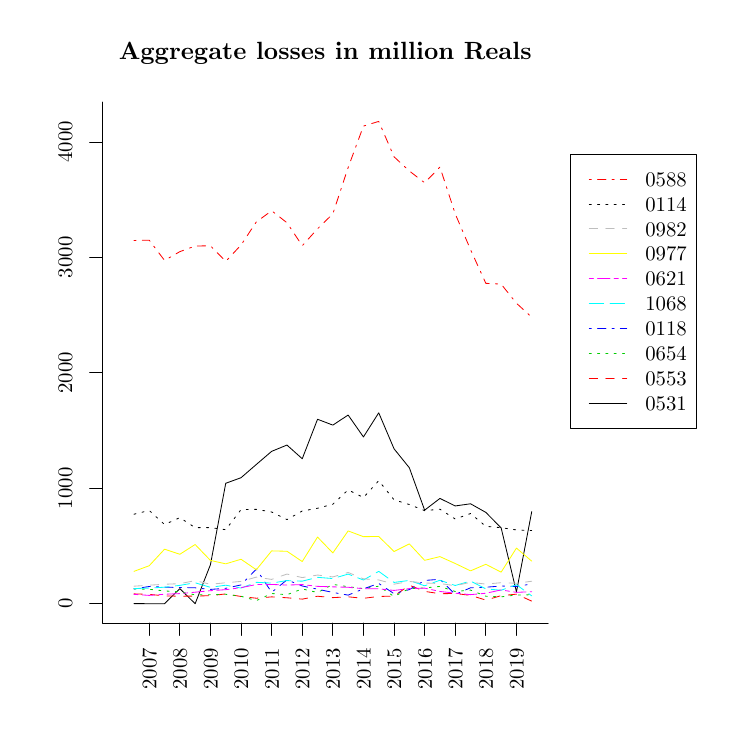}
		\caption{Evolution of the aggregate losses over time for different business lines using the SES data set, expressed in million Brazilian Reals, where the business lines are identified by their SUSEP code (see Table \ref{Table1}). }
		\label{Figure1}
	\end{center}
\end{figure}

\begin{figure}[t]
	\begin{center}
		\includegraphics{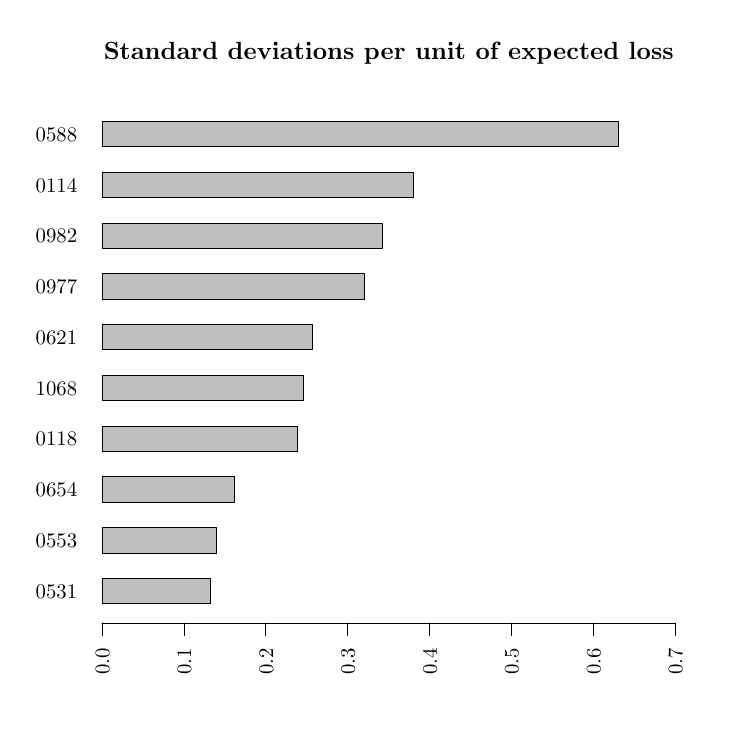}
		\caption{Ratio $\psi=\frac{\sigma}{\pi}$, i.e. the standard deviation per unit of expected loss, for all business lines, where the business lines are identified by their SUSEP code (see Table \ref{Table1}).}
		\label{Figure2}
	\end{center}
\end{figure}

The evolution of the aggregate losses over time are displayed on Figure \ref{Figure1}, and the corresponding ratios $\psi$ are displayed on Figure \ref{Figure2}. The minimum value of pairwise correlations is $-0.34$, and is found between comprehensive residential insurance (0114) and carrier insurance (0654). The maximum correlation is $0.86$ between optional auto liability (0553) and auto property damage (0531). The high positive correlation for the latter pair complies with the intuition.


Table \ref{Table2} contains the output of the analysis. Among the $45$ possible pairs, $34$ pairs satisfy the condition $b\rho<1$. This means that a large majority of $75.55\%$ of the possible pairs have a positive competitiveness region, and hence, can be priced jointly. Two business lines can be priced jointly with any of the other business lines included in the analysis, namely, insurance for comprehensive residential insurance (0114) and lender insurance (0977). Commercial multiple peril insurance (0118) can be priced jointly with any other business line, except motor third party liability (0588). The latter has the lowest number of matches, where it can be priced jointly with only three among the nine business lines. Specifically, motor third party liability (0588) can be priced jointly only with comprehensive residential insurance (0114), passenger private auto (0982), or lender insurance (0977).

The four non-commercial auto insurance business lines are all in the top five lines in terms of total collected premiums. The top two business lines from this category are auto property damage (0531) and optional auto liability (0553), and they are also the least risky ones. These two lines have a positive competitiveness region, which is rather surprising given that they also exhibit the highest pairwise correlation of 0.86. The other two lines are motor third party liability (0588) and private passengers auto (0982), and are the most risky ones. These two lines have a positive competitiveness region with a relatively low pairwise correlation of 0.18. Other combinations involving these four lines do not have a competitiveness region. This means that besides the two pairs reported above, one line would necessarily subsidize the other if they were to be priced jointly. 

The conclusion from this analysis is that there is room for insurance companies to price many business lines jointly, even when the pairwise correlation is high. Some pairs of business lines in related market segments, especially in auto insurance, do not have a competitiveness region. Nevertheless, candidate business lines for joint pricing do not necessarily need to be from the same market segment. In fact, it may be argued that unrelated lines are more likely to have a low correlation. But even for related lines with strong positive correlation, the competitiveness region may still exist, which is the case of auto property damage (0531) and optional auto liability (0553). 

Another relevant point to recall is that the decision on the pricing strategy should be supplemented by information about the price elasticity of demand in each line. To illustrate, consider the case of compulsory motor third party liability (0588) and optional auto liability (0553). The latter pays a benefit in excess of the cover provided by the former. Insurers are likely to be active on both of these business lines, but the analysis suggests that there exists no competitiveness region between them. As a consequence, pricing them jointly implies that the least risky business line (i.e. optional auto liability) would subsidize the riskier one (i.e. compulsory auto liability). In this case, and based on the analysis from Section \ref{Sec:Unconditional}, the insurer could still price them jointly if two conditions are satisfied. The first condition is that the compulsory line should be relatively elastic. The second condition is on the reaction of policyholders on the optional line, and this condition is revealed from the relative demand of this line compared to the compulsory cover. Note that it is not unrealistic to infer that the compulsory cover is likely to have the highest demand among the two, and hence, that the demand for the compulsory cover may be such that $w^d<w_{ct}$. If this is the case, then on top of the high elasticity on the compulsory business line, joint pricing would also require a high elasticity on the optional one.

\begin{table}[!h]
	\centering
	{\setlength{\extrarowheight}{5pt}
		\begin{tabular}{>{\centering}m{0.7cm}  >{\centering}m{0.7cm} >{\centering}m{0.7cm}  >{\centering}m{0.7cm} >{\centering}m{0.7cm}  >{\centering}m{0.7cm} >{\centering}m{0.7cm}  >{\centering}m{0.7cm} >{\centering}m{0.7cm}  >{\centering}m{0.7cm} >{\centering\arraybackslash}m{0.7cm} }
			\toprule 
			&      0588&      0114&      0982&      0977&      0621&      1068&      0118&      0654&      0553&      0531\\
			0588& $\bullet$&         +&         +&         +&         --&         --&         --&         --&         --&         --\\
			0114&         +& $\bullet$&         +&         +&         +&         +&         +&         +&         +&         +\\
			0982&         +&         +& $\bullet$&         +&         +&         +&         +&         +&         --&         --\\
			0977&         +&         +&         +& $\bullet$&         +&         +&         +&         +&         +&         +\\
			0621&         --&         +&         +&         +& $\bullet$&         +&         +&         +&         --&         --\\
			1068&         --&         +&         +&         +&         +& $\bullet$&         +&         --&         +&         +\\
			0118&         --&         +&         +&         +&         +&         +& $\bullet$&         +&         +&         +\\
			0654&         --&         +&         +&         +&         +&         --&         +& $\bullet$&         +&         +\\
			0553&         --&         +&         --&         +&         --&         +&         +&         +& $\bullet$&         +\\
			0531&         --&         +&         --&         +&         --&         +&         +&         +&         +& $\bullet$\\
			\bottomrule	
	\end{tabular}}
	\caption{Assessment of the condition $b\rho<1$ for all pairs of business lines, identified by their SUSEP code (see Table \ref{Table1}). The symbol (+) indicates that there exists a competitiveness region for the corresponding pair (i.e. the condition $b\rho<1$ is satisfied). The symbol (--) indicates that there is no competitiveness region for the corresponding pair (i.e. the condition $b\rho<1$ is not satisfied).}	\label{Table2}
\end{table}

\subsection{Illustration of the competitiveness region}
Two pairs of business lines with $\psi_A\leq \psi_B$ from the SES data set are used to illustrate the competitiveness region. For the first pair, business line $A$ corresponds to optional auto liability (0553) and business line $B$ corresponds to motor third party liability (1198). This pair is chosen because it has the highest value of $b\rho$, with $b\rho\approx0.49$. For the second pair, business line $A$ corresponds to mortgage insurance (1068) and business line $B$ corresponds to commercial multiple peril insurance (0118). This pair is selected because it has the lowest difference $\psi_B-\psi_A \approx 0.0069$, and in addition, satisfies $b\rho<1$, with $b\rho \approx 0.31$. 

\begin{figure}[t]
	\begin{center}
		\includegraphics[width=\textwidth]{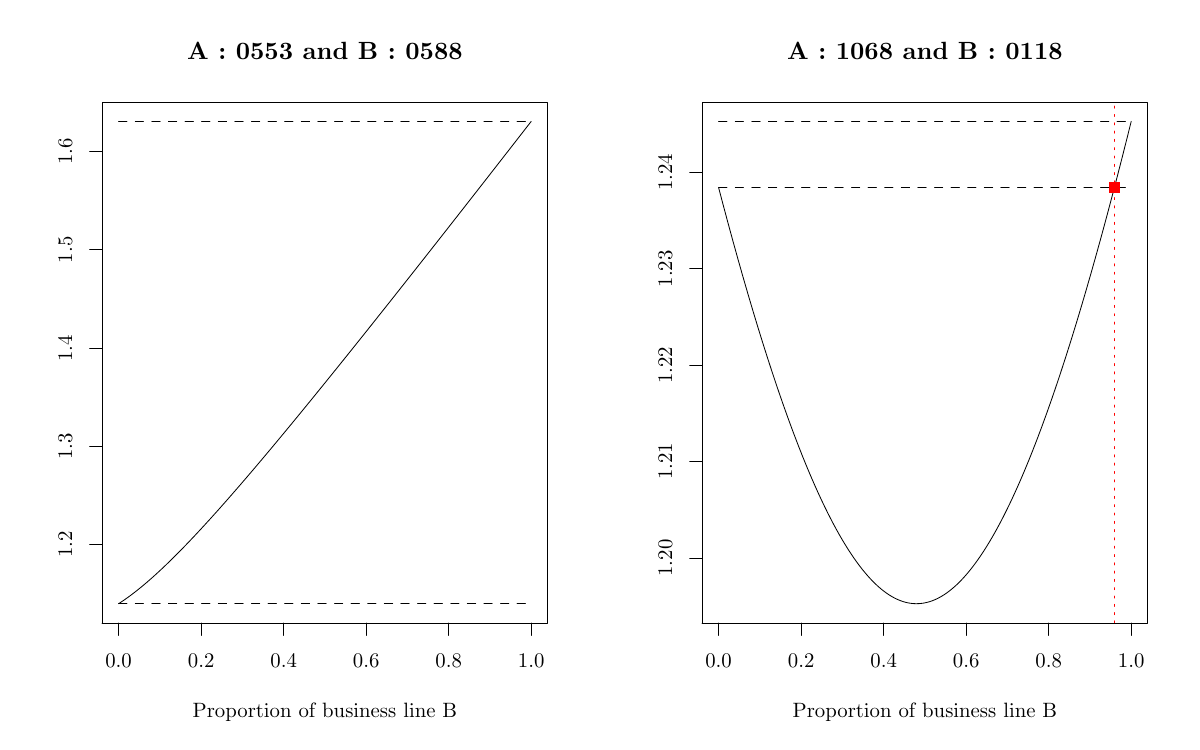}
		\caption{Loaded premiums under joint pricing per unit of expected benefit in function of the proportion of contracts in business line $B$, i.e. $1+\psi(n)$, for two selected pairs of business lines. The dashed horizontal lines are the stand-alone prices $1+\psi_A$ and $1+\psi_B$. The dashed vertical line on the right panel is the critical threshold.}
		\label{Figure3}
	\end{center}
\end{figure}

For each pair, the function $\psi(n)$ from \eqref{Eq4-3} is determined. Recall that this function corresponds to the required joint loading conditionally on the proportion of contracts in business line $B$. Figure \ref{Figure3} displays the loaded premiums under joint pricing per unit of expected benefit, i.e. $1+\psi(n)$. The dashed horizontal lines are the stand-alone prices $1+\psi_A$ and $1+\psi_B$.

The left panel of Figure \ref{Figure3} corresponds to the case where the competitiveness region does not exist. Moreover, due to the fact that condition $b\rho<1$ is not satisfied and that $\psi_A<\psi_B$, business line $A$ subsidizes business line $B$. Indeed, the lowest possible price for business line $A$ is its stand-alone price, whereas any proportion $n$ leads to a lower premium for business line $B$ under joint pricing. 

The right panel of Figure \ref{Figure3} illustrates a favorable situation for two business lines to be priced jointly. The existence of the competitiveness region is ensured by $b\rho<1$, whereas the critical threshold $n_{\textit{ct}}\approx 1$ is ensured by the fact that $\psi_A\approx\psi_B$. The remaining task for the insurer is to control the proportion $n$ in order for the actual loading to be above the required one. Alternatively, the insurer can gain more flexibility in terms of portfolio monitoring if the commercial premiums under joint pricing are set by taking into account information on market structure and policyholders' behavior in an appropriate way.

\section{Concluding remarks}
\label{Conclusions}
This paper investigates how diversification effects can be incorporated into prices of insurance policies. The central question is whether insurers active on two business lines with offsetting liabilities should price them jointly or separately. The paper analyzes the required risk premiums, and shows that a joint pricer has a potential for competitiveness, but this potential comes with the burden of portfolio monitoring. Further, the number of underwritten policies is modeled in function of market structure and policyholders' behavior. This allows to obtain conditions on the reaction factors of policyholders ensuring the optimality of joint pricing over stand-alone pricing, where the decision criterion is the total collected premiums. The analysis is illustrated for a portfolio of term annuities and term assurances, and is supplemented by an empirical exploration of aggregate losses using data of non-life business lines from the Brazilian insurance market.

This work allows to answer some questions related to joint pricing, and paves the path for potential future research topics. A particularly interesting research endeavor would consist in studying joint pricing in a multivariate framework, which would allow insurers to select and price a larger number of business lines jointly. Further, a natural step forward would consist in extending the present analysis to a dynamic setting, as in \cite{Taylor,Emms} or \cite{PantelousPassalidou}. On the one hand, this would allow to take into account offsetting effects over time within and across business lines. On the other hand, a dynamic setting would also allow to incorporate some features of the property-casualty insurance industry such as the underwriting cycle \citep{Venezian1985,CumminsHarringtonKlein1991,MeierOutreville2006}.

Another relevant issue that could be investigated is whether the main conclusions of the paper would change by taking into account some distributional features beyond the second moments and linear correlations. For instance, this could be necessary for some long-tail business lines exhibiting skewness, and pairs of business lines exhibiting asymmetric dependence structures. Nevertheless, it is worth noting that in the present paper, the analysis in Section \ref{Illustration} shows that the MSD can be a good substitute for the value-at-risk in applications involving annuities and assurances. Unreported tests show that it can also be a good substitute for the conditional value-at-risk in the same context. Therefore, the main conclusions of the paper remain valid for this type of contracts under mortality models such as the Li-Lee model, which are also used in industry.

It should be noted that despite the potential benefits of diversification across business lines in the pricing, regulators can restrict firms from operating on some specific business lines simultaneously. For instance, unlike in other Latin American countries, Brazilian insurers providing health or credit export insurance must be specialized. Further, it is not allowed for a Brazilian life insurance company providing open pension funds to sell non-life policies \citep{OECD}. One rationale behind these restrictions is that some lines are deemed too risky, and the insurer's default on one of those lines would impact customers' welfare from other lines \citep{Jaffee,Ibragimovetal2018}. Another typical such restriction requires separating life and non-life insurance activities; see e.g. EIOPA Single Rulebook for European insurers, and \cite{OECD} for Latin American countries. Nevertheless, firms are allowed to operate simultaneously on a substantial number of lines, and the analysis in the present paper remains relevant in many cases.

Joint pricing may also refer to bundling; see e.g. \cite{Bhargava} and \cite{GokgurKarabati}. In the insurance industry, bundling is also subject to legal constraints in some countries. For instance, Brazil, Chile and Latvia have strong restrictions on bundled insurance products in general, and in the US, home and auto insurance cannot be bundled \citep{OECD2020}. Joint pricing is however not confined to bundling. Bundling consists in selling multiple items with a related function to a single customer, whereas in the present context, joint pricing is implemented at portfolio level for potentially two different types of customers. For instance, annuities would be purchased by elderly seeking income protection, whereas assurances would be purchased by active policyholders in order to protect their dependents from a loss of income in case of death.
\bibliographystyle{agsm}
\bibliography{References}

\appendix
\section{Proofs}
\renewcommand{\theequation}{\thesection.\arabic{equation}}

\subsection{Proof of Lemma \ref{Lemma-1}}
\label{Lemma-1-Proof}
The Chain rule leads to $\frac{\partial\Psi(n)}{\partial n}=\frac{\partial\Psi(n)}{\partial \tilde{n}}\frac{\partial \tilde{n}}{\partial n}$. Since $\frac{\partial \tilde{n}}{\partial n}>0$, an optimum of $\psi(n)$ is such that $\frac{\partial\Psi(n)}{\partial \tilde{n}}=0$. The derivative of $\psi(n)$ with respect to $\tilde{n}$ is given by:
$$\frac{\partial\psi(n)}{\partial \tilde{n}}=\psi_A\frac{\lambda_1 \tilde{n} - \lambda_2}{\left(\lambda_1\tilde{n}^2 - 2\lambda_2\tilde{n} + 1\right)^{\frac{1}{2}}},$$
which is equal to $0$ for $\tilde{n} = \frac{\lambda_2}{\lambda_1}$, or equivalently, for $n=n^{\text{opt}}=\frac{\lambda_{2}\pi_A}{\lambda_{2}\pi_A+\left(\lambda_1-\lambda_2\right)\pi_B}$. 

Recall that $b=1$. The argument $n^{\text{opt}}$ of this optimum is unique. For $\lambda_2=0$ (i.e. $b\rho=1$), this argument is equal to $1$, whereas for $\lambda_2=\lambda_1$ (i.e. $b=\rho$, and hence, for $b=\rho=1$), it is equal to $1$. 

Evaluated at the optimum $n^{\text{opt}}$, the function $\psi(n)$ is equal to:
$$ \psi(n^{\text{opt}}) =\psi_A\sqrt{\frac{\lambda_1-\lambda_2^2}{\lambda_1}},$$
where $\lambda_2^2=1+b^2\rho^2-2b\rho$, which leads to $\lambda_1-\lambda_2^2=b^2(1-\rho^2)\geq 0$. Since $\lambda_2^2>0$, it follows that $\frac{\lambda_1-\lambda_2^2}{\lambda_1}>1$, and hence $\psi(n^{\text{opt}})<\psi_A$. This implies that $n^{\text{opt}}$ leads to a minimum, where the notation $n^{\text{opt}}=n^{\min}$ is used.

The remaining task is to determine conditions under which the argument $n^{\min}$ is within the interval $(0,1)$. From the discussion above, it follows that $\lambda_2=0$ is equivalent with $b\rho=1$, and hence $n^{\min}=0\notin(0,1)$. For $\lambda_2\neq0$, it follows that $n^{\min}\in(0,1)$ provided $1 + \frac{\pi_B}{\pi_A}\left(\frac{\lambda_1}{\lambda_2}-1\right)>0$ and $\frac{\lambda_1}{\lambda_2}>1$. Note that the former condition is always satisfied when the latter condition is satisfied. Thus, $n^{\min}\in(0,1)$ if and only if $\frac{\lambda_1}{\lambda_2}>1$, where $\lambda_1>0$ always holds. For $\lambda_2<0$, this condition is never satisfied. For $\lambda_2>0$, the condition is satisfied for $\lambda_1>\lambda_2$, which is always the case for $\rho<1\leq b$. Since $\lambda_2>0$ is equivalent with $b\rho<1$, it follows that $n^{\min}\in(0,1)$ if and only if $b\rho<1$.

\subsection{Proof of Lemma \ref{Lemma-2}}\label{Lemma-2-Proof}
The proof consists in solving the equation $P^{\textit{jp}}_{A}(n)=P^{\textit{sa}}_A$. By setting $\psi^{\star}=\psi_A$ in \eqref{Eq-21} and \eqref{Eq-22}, it directly follows that $n^{\star}_l =\tilde{n}^{\star}_l=0$ and $\tilde{n}^{\star}_u  = 2\frac{\lambda_2}{\lambda_1}$, where the latter can be transformed to find the expression of $n_{\textit{ct}}$. Using the convexity of $\psi(n)$ ends the proof.

\subsection{Proof of Theorem \ref{Theorem-1}}\label{Theorem-1-Proof}
The difference $\mathcal{D}_{\textit{ptf}}$ can be expressed in two forms, namely, as a quadratic function of $1+\psi^\star$:
\begin{equation}\label{Eq1}\mathcal{D}_{\textit{ptf}} = -\theta_1 \left(1+\psi^{\star}\right)^2 + \theta_2 \left(1+\psi^{\star}\right) - \theta_3,\end{equation}
where
$$\begin{array}{ll}
	\theta_1 &= q_A\frac{N^T_A}{k_A}\pi_A\frac{k_A-1}{k_A}\frac{1}{1+\psi_A}+q_B\frac{N^T_B}{k_B}\pi_B\frac{k_B-1}{k_B}\frac{1}{1+\psi_B},\\
	\theta_2 & =\left(q_A\frac{N^T_A}{k_A}\pi_A\frac{k_A-1}{k_A} + q_B\frac{N^T_B}{k_B}\pi_B\frac{k_B-1}{k_B}\right) +  \left(\frac{N^T_A}{k_A}\pi_A + \frac{N^T_B}{k_B}\pi_B\right),\\
	\theta_3 &= \frac{N^T_A}{k_A}\pi_A(1+\psi_A)+\frac{N^T_B}{k_B}\pi_B(1+\psi_B),\end{array}$$
or, equivalently, as a linear combination of the reaction factors $q_A$ and $q_B$:
\begin{equation}\label{Eq2} \begin{array}{ll}\frac{\mathcal{D}_{\textit{ptf}}}{\frac{N_A}{k_A}\pi_A + \frac{N_B}{k_B}\pi_B}=& q_A\frac{k_A-1}{k_A}\frac{1+\psi^{\star}}{1+\psi_A}(1-\bar{n})(\psi_A-\psi^{\star})+q_B\frac{k_B-1}{k_B}\frac{1+\psi^{\star}}{1+\psi_B}\bar{n}(\psi_B-\psi^{\star})\\&-\left((1-\bar{n})\left(\psi_A-\psi^{\star}\right) + \bar{n}(\psi_B-\psi^{\star})\right),\end{array}
\end{equation}
where
$$\bar{n} = \frac{\frac{N_B}{k_B}\pi_B}{\frac{N_A}{k_A}\pi_A + \frac{N_B}{k_B}\pi_B}.$$

The proof uses the results of Lemma \ref{Lemma-2} on the critical threshold. In particular, $\psi^{\star}\in[\psi^{\min},\psi_A)$ for $n^{\star}\in(0,n_{\textit{ct}})$, $\psi^{\star}\in(\psi_A,\psi_B)$ for $n^{\star}\in(n_{\textit{ct}},1)$ and $\psi^{\star}=\psi_A$ for $n^{\star}=n_{\textit{ct}}$, where $\psi^{\min}$ is given in \eqref{Psimin}, and $n_{\textit{ct}}$ has the following expression:
$$n_{\textit{ct}}=\frac{2 \lambda_2\pi_A}{2 \lambda_2\pi_A + (\lambda_1-2 \lambda_2)\pi_B}.$$
The critical threshold $n_{\textit{ct}}$ in the proportion of underwritten contracts $B$ is equivalent to a critical threshold $w_{\textit{ct}}$ in the proportion of market demand for contract $B$, such that:
$$n_{\textit{ct}}=\frac{w_{\textit{ct}}\frac{1}{k_B}\left(1+\frac{k_B-1}{k_B}\frac{\psi_B-\psi_A}{1+\psi_B}q_B\right)}{(1-w_{\textit{ct}})\frac{1}{k_A}+w_{\textit{ct}}\frac{1}{k_B}\left(1+\frac{k_B-1}{k_B}\frac{\psi_B-\psi_A}{1+\psi_B}q_B\right)}.$$
Hence, $w_{\textit{ct}}$ is given by:
$$w_{\textit{ct}}=\frac{n_{\textit{ct}}}{n_{\textit{ct}} + (1-n_{\textit{ct}})\frac{k_A}{k_B}\left(1+\frac{k_B-1}{k_B}\frac{\psi_B-\psi_A}{1+\psi_B}q_B\right)}.$$
Note that $w^d=0$ and $w^{d}=w_{\textit{ct}}$ are the only proportions of demand such that $\psi^{\star}=\psi_A$. Moreover, $n^{\star}=1$ for $w^d=1$, and hence $\psi^{\star}=\psi_B$. This confirms that $\psi^{\star}\in[\psi^{\min},\psi_A)$ for $w^d<w_{\textit{ct}}$, $\psi^{\star}\in(\psi_A,\psi_B)$ for $w^d>w_{\textit{ct}}$, and $\psi^{\star}=\psi_A$ for $w^d=w_{\textit{ct}}$.

\subsubsection{Part 1: $\mathcal{D}_{\textit{ptf}}>0$}
The proof begins with the first part of the theorem and the conditions such that $\mathcal{D}_{\textit{ptf}}>0$, which means that it is more advantageous to price the contracts jointly. The proof of this first part treats the cases $b>1$ and $b=1$ separately.

\textbf{$\star$ Case 1: $b>1$}

First, suppose that $b>1$, which is equivalent with $\psi_B>\psi_A$. Since $\theta_1>0$, the difference $\mathcal{D}_{\textit{ptf}}$ is a concave function of $\psi^{\star}$, with $\underset{\psi^{\star}\rightarrow \pm \infty}{\lim}\mathcal{D}_{\textit{ptf}}(\psi^{\star})=-\infty$. Further, $\psi^{\star}\in[\psi^{\min},\psi_A)$ for $w^d<w_{\textit{ct}}$, and $\psi^{\star}\in(\psi_A,\psi_B)$ for $w^d>w_{\textit{ct}}$, as well as $\psi^{\star}=\psi_A$ for $w^d=w_{\textit{ct}}$. In the first two cases, it is sufficient to show that the concave function $\psi\mapsto -\theta_1(1+\psi)^2+\theta_2(1+\psi)-\theta_3$, evaluated at the end points of the domain of $\psi^{\star}$, is positive.

Thus, for $w_d <w_{\textit{ct}}$, the difference $\mathcal{D}_{\textit{ptf}}$ is always positive if the inequalities:
$$\begin{array}{ll}
	\theta_1(1+\psi^{\min})^2 - \theta_2(1+\psi^{\min}) + \theta_3 & <0,\\
	\theta_1(1+\psi_A)^2 - \theta_2(1+\psi_A) + \theta_3 & <0,
\end{array}$$
are both satisfied. On the other hand, for $w_d>w_{\textit{ct}}$, the difference $\mathcal{D}_{\textit{ptf}}$ is always positive if the inequalities:
$$\begin{array}{ll}
	\theta_1(1+\psi_A)^2 - \theta_2(1+\psi_A) + \theta_3 & <0,\\
	\theta_1(1+\psi_B)^2 - \theta_2(1+\psi_B) + \theta_3 & <0,
\end{array}$$
are both satisfied. This means that $\mathcal{D}_{\textit{ptf}}>0$ if $$\theta_1(1+\psi_A)^2 - \theta_2(1+\psi_A) + \theta_3 <0,$$ and:
$$\left\{\begin{array}{lll}\theta_1(1+\psi^{\min})^2 - \theta_2(1+\psi^{\min}) + \theta_3 & <0 & \qquad \text{for } w_d<w_{\textit{ct}},\\
	\theta_1(1+\psi_B)^2 - \theta_2(1+\psi_B) + \theta_3 & <0 & \qquad \text{for } w_d>w_{\textit{ct}}. \end{array} \right.$$
Inequality $\theta_1(1+\psi_A)^2 - \theta_2(1+\psi_A) + \theta_3 <0$ is equivalent with:
$$q_B >\frac{k_B}{k_B-1}\frac{1+\psi_B}{1+\psi_A}.$$
Inequality $\theta_1(1+\psi_B)^2 - \theta_2(1+\psi_B) + \theta_3<0$ is equivalent with:
$$q_A  <\frac{k_A}{k_A-1}\frac{1+\psi_A}{1+\psi_B}.$$
Inequality $\theta_1(1+\psi^{\min})^2 - \theta_2(1+\psi^{\min}) + \theta_3 <0$ is equivalent with:
\begin{equation}\label{Eq3x}(1-\eta)q_A\frac{k_A-1}{k_A}\frac{1+\psi^{\min}}{1+\psi_A}+\eta q_B\frac{k_B-1}{k_B}\frac{1+\psi^{\min}}{1+\psi_B}>1,\end{equation}
where
$$\eta = \frac{\frac{N^T_B}{k_B}\pi_B(\psi_B-\psi^{\min})}{\frac{N^T_A}{k_A}\pi_A(\psi_A-\psi^{\min})+\frac{N^T_B}{k_B}\pi_B(\psi_B-\psi^{\min})}.$$
Consider now the case where $w^d=w_{\textit{ct}}$, which is equivalent with $\psi^{\star}=\psi_A$. From \eqref{Eq2}, $\mathcal{D}_{\textit{ptf}}>0$ is equivalent with:
\begin{equation}\label{Eq3}q_A\frac{k_A-1}{k_A}\frac{1+\psi^{\star}}{1+\psi_A}(1-\bar{n})(\psi_A-\psi^{\star})+q_B\frac{k_B-1}{k_B}\frac{1+\psi^{\star}}{1+\psi_B}\bar{n}(\psi_B-\psi^{\star})>(1-\bar{n})\left(\psi_A-\psi^{\star}\right) + \bar{n}(\psi_B-\psi^{\star}),\end{equation}
and hence, for $\psi^{\star}=\psi_A$, it follows that inequality \eqref{Eq3} reduces to:
$$q_B>\frac{k_B}{k_B-1}\frac{1+\psi_B}{1+\psi_A},$$
which ends the proof for the case where $b>1$.

\textbf{$\star$ Case 2: $b=1$}

For $b=1$, which means that $\psi_A=\psi_B$, inequality \eqref{Eq3} becomes:
$$q_A\frac{k_A-1}{k_A}\frac{1+\psi^{\star}}{1+\psi_A}(1-\bar{n})+q_B\frac{k_B-1}{k_B}\frac{1+\psi^{\star}}{1+\psi_B}\bar{n}>1.$$
Moreover, since $\psi^{\star}\geq\psi^{\min}$, then a sufficient condition for $\mathcal{D}_{\textit{ptf}}>0$ is given by:
$$q_A\frac{k_A-1}{k_A}(1-\bar{n})+q_B\frac{k_B-1}{k_B}\bar{n}>\frac{1+\psi_B}{1+\psi^{\min}},$$
which is, by noting that $\eta=\bar{n}$ for $\psi_A=\psi_B$, a simplified version of inequality \eqref{Eq3x}.

\subsubsection{Part 2: $\mathcal{D}_{\textit{ptf}}<0$}

The second part of the proof consists in deriving sufficient conditions such that $\mathcal{D}_{\textit{ptf}}<0$, i.e. when the insurer is better off with stand-alone pricing.
From inequality \eqref{Eq3}, $\mathcal{D}_{\textit{ptf}}$ is always negative for:
\begin{equation}\label{Eq4}q_B\frac{k_B-1}{k_B}\frac{1+\psi^{\star}}{1+\psi_B}<1,\end{equation}
and
\begin{equation}\label{Eq5}q_A\frac{k_A-1}{k_A}\frac{1+\psi^{\star}}{1+\psi_A}(\psi_A-\psi^{\star})<\psi_A-\psi^{\star},\end{equation}
where only one of the two inequalities needs to be strict.

Recall that $\psi^{\star}\in[\psi^{\min},\psi_A)$ for $w^d<w_{\textit{ct}}$, that $\psi^{\star}\in(\psi_A,\psi_B)$ for $w^d>w_{\textit{ct}}$, and that $\psi^{\star}=\psi_A$ for $w^d=w_{\textit{ct}}$. Thus, inequality \eqref{Eq4} is satisfied when:
$$\left\{ \begin{array}{ll} q_B<\frac{k_B}{k_B-1}\frac{1+\psi_B}{1+\psi_A},& \qquad \text{for } w^d<w_{\textit{ct}},\\ q_B<\frac{k_B}{k_B-1},& \qquad\text{for } w^d\geq w_{\textit{ct}},\end{array}\right.$$
or, simply for $q_B<\frac{k_B}{k_B-1}$. Moreover, inequality \eqref{Eq5} is satisfied when:
$$\left\{ \begin{array}{ll} q_A<\frac{k_A}{k_A-1},& \qquad \text{for } w^d<w_{\textit{ct}},\\ q_A>\frac{k_A}{k_A-1},& \qquad\text{for } w^d> w_{\textit{ct}}.\end{array}\right.$$
Note that only inequality \eqref{Eq4} is sufficient for $w^d=w_{\textit{ct}}$, and that these conditions are valid for $b=1$.

\subsection{Proof of Corollary \ref{Corollary}}\label{Corollary-proof}
Since $k_B$ and $k_A$ are at least equal to $2$, it is straightforward that $\frac{k_A}{k_A-1}$ and $\frac{k_B}{k_B-1}$ are both in $[2,1)$. Moreover, since $\psi_A>\psi^{\min}$, which implies that as sufficient condition for inequality \eqref{EqT1-1} to be satisfied is given by:
$$q_B>2\frac{1+\psi_B}{1+\psi^{\min}}.$$
Provided this latter inequality is satisfied, then for $w^d<w_{\textit{ct}}$, inequality
$$q_A>2\frac{1+\psi_A}{1+\psi^{\min}}$$
ensures that $(1-\eta)q_A\frac{k_A-1}{k_A}\frac{1+\psi^{\min}}{1+\psi_A}+\eta q_B\frac{k_B-1}{k_B}\frac{1+\psi^{\min}}{1+\psi_{B}}>1$. The remaining inequalities can be derived by investigating the limits of $\frac{k_A}{k_A-1}$ and $\frac{k_B}{k_B-1}$ for either $k=2$ or $k\rightarrow \infty$.

\end{document}